%% file: main.tex
\documentclass[a4paper,11pt]{article}
\usepackage{jheppub} 
\usepackage{lineno}
\usepackage{tikz-feynman}
\usepackage{slashed}
\usepackage[T1]{fontenc} 
\usepackage[utf8]{inputenc}
\usepackage{arydshln}
\usepackage{pdflscape}
\usepackage{amssymb}
\usepackage{tikz}
\usepackage{subfigure}
\usepackage{multirow}

\usepackage{tabularray}
\newcommand{\SPtt}[1]{\bar{u}_t #1 v_{\bar{t}}}
\newcommand{\SPqq}[1]{\bar{v}_{\bar{q}} #1 u_q}
\newcommand{\Opp}[2]{\OO_{\sss #1}^{\sss #2}}

\newcommand{\tvp}{\widetilde{\varphi}}

\newcommand{\OO}{\ensuremath{\mathcal{O}}}

\renewcommand{\phi}{\ensuremath{\varphi}}

\newcommand{\sss}{\scriptscriptstyle}
\newcommand{\bpm}{\begin{pmatrix}}      
\newcommand{\epm}{\end{pmatrix}}

\newcommand{\Op}[1]{\OO_{\sss #1}}

\definecolor{lightgray}{rgb}{0.83, 0.83, 0.83}
\definecolor{lightpurp}{rgb}{0.901,0.796,0.882}

\input{diagrams}

\title{\boldmath Going beyond Top EFT}

\author[a]{Andr\'e Lessa}
\author[b]{and Ver\'onica Sanz}
\affiliation[a]{Centro de Ciências Naturais e Humanas, Universidade Federal do ABC,
Santo André, 09210-580 SP, Brazil}
\affiliation[b]{Instituto de F\'isica Corpuscular (IFIC), Universidad de Valencia-CSIC, E-46980 Valencia, Spain}

\emailAdd{andre.lessa@ufabc.edu.br}
\emailAdd{veronica.sanz@uv.es}

\abstract{
 We present a new way to interpret Top Standard Model measurements going beyond the SMEFT framework. Instead of the usual paradigm in Top EFT, where the main effects come from tails in momenta distributions, we propose an interpretation in terms of new physics which only shows up at loop-level. The effects of these new states, which can be lighter than required within the SMEFT, appear as distinctive structures at high momenta, but may be suppressed at the tails of distributions. As an illustration of this phenomena, we present the explicit case of a UV model with a $\mathcal{Z}_2$ symmetry, including a Dark Matter candidate and a top-partner. This simple UV model reproduces the main features of this class of signatures, particularly a momentum-dependent form factor with more structure than the SMEFT. As the new states can be lighter than in SMEFT, we explore the interplay between the reinterpretation of direct searches for colored states and Dark Matter, and Top measurements, made by ATLAS and CMS in the differential $t \bar t$ final state. We also compare our method with what one would expect using the SMEFT reinterpretation, finding that using the full loop information provides a better discriminating power.}

\begin{document}
\maketitle
\flushbottom

\section{Introduction}
\label{sec:intro}

At the LHC, searches for new phenomena in the Standard Model Effective Field Theory (SMEFT) framework are now commonplace, see e.g. Refs.~\cite{CMS:2023ixc,CMS:2022hjj,CMS:2022uox,CMS:2020gsy,CMS:2021cxr,ATLAS:2022xyx,ATLAS:2021jgw,ATLAS:2023jma} for recent experimental results. They provide a way to re-interpret Standard Model (SM) measurements which exploits their full kinematic range and can guide combinations of different channels. The most striking signatures of SMEFT show up at tails in energy-momentum distributions~\cite{Ellis:2014dva}, extreme kinematic regions where the SM contribution is scarce and the new phenomena more visible. A similar story can be told for light axion-like particles (ALPs), whose derivative couplings also induce prominent effects in the tails of distributions from SM measurements~\cite{Gavela:2019cmq,Folgado:2020utn}. 

On the other hand, direct searches for new physics are based on on-shell production of the new states, which can then decay, interact with the detector or escape detection. Direct searches can be based on signatures with very low SM background and/or searches for excesses 
in specific channels and phase space regions.
For instance, in the case of bump-hunt searches, one would scan for deviations from a smooth SM background in a range of resonance masses.
Nowadays, these searches are sensitive to very high masses (well above TeV) for traditional channels  (e.g., dijet or dilepton) and the LHC experimental collaborations are continuously broadening the coverage for possible final states. Despite their impressive sensitivity, resonance  searches have an intrinsic limitation: they make sense for narrow states, with widths ($\Gamma$) much smaller than their mass ($m$),  and typically their performance quickly stops at $\Gamma/m \gtrsim$ 0.3, see e.g. Refs.~\cite{CMS:2021ctt,CMS:2019gwf} for recent experimental analyses with variable widths~\footnote{A theoretical proposal to broaden the scope of these searches has been presented in Ref.~\cite{Chivukula:2017lyk}.}.
In addition, due to trigger requirements, soft final states which can appear in compressed Beyond the Standard Model (BSM) scenarios can also be very challenging for direct searches, which must rely on initial or final state radiation for triggering.

Somewhere in between the SMEFT and on-shell paradigms lies the proposal of this paper, namely the exploration of scenarios that are just beyond the reach of direct searches, but are not correctly described by the Effective Field Theory (EFT) limit. 
As we will show, these scenarios can be probed by SM measurements, but their signal can be very distinct from what would be expected from SMEFT.
In particular, if the new states contribute to SM measurements through loop diagrams, their effect in the differential distributions would be localized in a kinematic region, resembling a very broad bump. In the limit that the states running in the loop are very heavy, this localization would shift towards high invariant masses, reaching the SMEFT limit. 

As discussed in Ref.~\cite{Cepedello:2022pyx}, scenarios which contribute at loop-level at leading order are a good testing bed for the interplay between direct searches and indirect probes, as typically the new states can be lighter than in scenarios with tree-level contributions. And among the set of loop-induced UV models, those with a Dark Matter (DM) candidate are particularly interesting~\cite{Cepedello:2023yao}. Moreover, we will focus on scenarios with a special relation with the top sector, which will allow us to draw a comparison with the current efforts on the Top EFT searches and provide an alternative to those.
A similar approach was considered in Ref.~\cite{Drozd:2015kva}, but within the context of Higgs couplings.

\
The paper is structured as follows. In Sec.~\ref{sec:topeft} we review the main concepts in the Top sector of the SMEFT. In the section~\ref{sec:beyondtop}, we present a minimal loop-induced scenario with a DM particle and a heavy top-partner, discuss the analytical behaviour of their contribution to top observables, and explore the connection to the Top EFT. In Sec.~\ref{sec:lhc} the limits from direct searches are reviewed, as well as the limits from precise SM top observables ($m_{t \bar t}$ and $p_T$ distributions). Those direct and indirect probes are placed together and compared with the Top EFT limit in Sec.~\ref{sec:results}. Finally, in Sec.~\ref{sec:concls}, we conclude.
Auxiliary information concerning the loop calculation, matching to the EFT regime and the limit setting are given in Appendices~\ref{app:diagrams}, \ref{app:formFactors} and \ref{app:expectedLimits}.

The datasets used to obtain all the results presented here as well as additional details are available in the \href{https://doi.org/10.5281/zenodo.10277973}{Zenodo}~\cite{zenodoRepo} and \href{https://github.com/andlessa/SMStoEFT/tree/zenodo}{GitHub} repositories.

\section{Top EFT}
\label{sec:topeft}

One of the goals of this paper is to show a different way to interpret SM measurements, namely to search for new states which can be relatively light when compared to the LHC energy scale and only
contribute to higher dimensional operators at the loop level. To illustrate this point, we will show results from a UV extension of the SM with a singlet fermionic Dark Matter candidate and a scalar colored state, a partner of the right-handed top, $t_R$. For energy scales sufficiently smaller than the BSM masses, this scenario and its phenomenology can be matched to a reduced set of Top EFT operators.

The top EFT has been described in many works,  e.g.~\cite{Buckley:2015lku,Etesami:2017ufk,Maltoni:2019aot,Brivio:2019ius,Miralles:2021dyw,Kassabov:2023hbm}, and it is a subset of the SMEFT Lagrangian. In this paper we will use the Warsaw~\cite{Grzadkowski:2010es} convention to classify the independent operators.
Note that if we allowed for the most general flavour structure, we would find that there are 2499 different types of operators which contribute at dimension-six~\cite{Alonso:2013hga}, but this number is drastically reduced once we assume some type of flavour structure in the UV completions, an assumption well motivated by the obstinate absence of anomalies in flavour observables. In particular, when focusing on the top physics, it is common to consider a flavour $SU(3)^5$ symmetry, leading to a {\it top-specific} scenario. The details of this scenario, including a classification of the relevant operators and their limits from a global fit can be found in Ref.~\cite{Ellis:2020unq}. 
As we will show below, the particular UV model considered here induces a new interaction of two tops with gluons, 
$$\Op{tG}=(\bar Q \sigma^{\mu\nu} T^A t_R) \tvp\, G_{\mu\nu}^A, $$
where $\sigma^{\mu\nu} = \frac{i}{2} \left[\gamma^\mu, \gamma^\nu\right]$. This gluon-top coupling modifier will be accompanied by a  set 
of four-fermion interactions between the right-handed tops and the quark doublets (of any flavor), $q_{L}$:
\begin{equation*}
\Opp{tq}{(8)}=(\bar q_L \gamma_\mu T^A q_L)(\bar t_R \gamma^\mu T^A t_R),   
\end{equation*}
and the coupling between right-handed tops and right-handed quarks,
\begin{equation*}
\Opp{tt}{(8)}=(\bar t_R \gamma_\mu T^A t_R)(\bar t_R \gamma^\mu T^A t_R), \, \Opp{t(u/d)}{(8)}=(\bar t_R \gamma_\mu T^A t_R)((\bar u_R/\bar d_R) \gamma^\mu T^A (u_R/d_R)).
\end{equation*}

In Ref.~\cite{Ellis:2020unq}, we showed that the operator $\Op{tG}$ was mostly constrained by  Run 1 and Run 2 Higgs observables plus the TeVatron and LHC  $t \bar t$ datasets. Moreover, among the four-fermion operators, only the $\Opp{tX}{(8)}$ ($X$= $q$, $u$ or $d$) operators were constrained, predominantly by the top data and, to a lesser extent, by $t\bar t V$ measurements. The current limits on these four operators are shown in Table~\ref{tab:limits}, and one can see that the difference between the individual and marginalised limits is quite dramatic, particularly for the four-fermion operators.  The reason is that marginalised limits correspond to a global fit to many operators, beyond these shown here, which contribute to the same set of observables. Hence, the inclusion of more operators tend to weaken the limits for each operator. 

\begin{table}[t!]
    \centering
    \begin{tabular}{|c|c|c|}
    \hline 
         Operator & Individual fit (TeV$^{-2}$) & Marginalised fit (TeV$^{-2}$) \\
         \hline
         $\Op{tG}$& $-0.01^{+0.086}_{-0.1}$  & 0.36$^{+0.12}_{-0.6}$   \\
        $\Opp{tq}{(8)}$ & -0.4$^{+0.06}_{-0.85}$ & 5.$^{+2.2}_{-13}$    \\
        $\Opp{tu}{(8)}$ & -0.45$^{+0.23}_{-1.1}$  & 4.0 $^{+19}_{-11}$  \\
        $\Opp{td}{(8)}$ &-1.0$^{+0.38}_{-2.5}$ & -0.42$^{+11}_{-12}$ \\\hline
    \end{tabular}
    \caption{Limits on the relevant subset of Top operators at 95\% C.L., from Ref.~\cite{Ellis:2020unq}.  }
    \label{tab:limits}
\end{table}

\section{Beyond Top EFT}\label{sec:beyondtop}

The Top EFT operators discussed in the previous section are useful for describing new physics effects on energy scales well below the BSM masses. However, once we consider the TeV energies probed by the LHC and BSM particles with masses around 1~TeV, the validity of the EFT regime is not guaranteed.
In this case we need to go beyond the Top EFT and consider the UV extension of the SM.
Motivated by Dark Matter, we consider a BSM scenario with a $\mathcal{Z}_2$ parity, which ensures the stability of the DM candidate. An important consequence of this assumption is that it forbids linear couplings of the new states to two SM particles and the SMEFT operators are only induced at the one-loop level~\cite{Cepedello:2022pyx}.
In addition we will build in this model a special connection to the top sector, a possibility that has been partly explored in the context of DM relic abundance and collider phenomenology, see Refs.~\cite{Haisch:2015ioa,Plehn:2017bys,Delgado:2016umt,Garny:2018icg}.

\subsection{An explicit example: a UV extension with Dark Matter and a top partner}
\label{sec:uvmodel}

In order to incorporate the main features described above and be minimal, we consider the simple case of a scalar top partner ($\phi_T$), singlet under $SU(2)_L$, and a singlet fermion ($\chi$),
which is a Dark Matter candidate. Under the imposed $\mathcal{Z}_2$ symmetry the BSM fields are odd and the SM are even, so the renormalizable BSM lagrangian becomes:
\begin{equation}
    \mathcal{L}_{BSM} = \bar{\chi}\left( i \slashed{\partial} -\frac{1}{2} m_{\chi} \right) \chi + |D_\mu \phi_T|^2 - m_{T}^2 |\phi_T|^2 - \left( y_{\mathrm{DM}}\phi_T^\dagger \bar{\chi} t_R  + h.c. \right)
\end{equation}
with $m_T > m_{\chi}$, so the DM candidate is stable.
The only viable decay channel for $\phi_T$ is $\phi_T \to \chi + t$, where the top is off-shell if $\Delta M = m_T - m_{\chi} < m_t$.
Note that the interactions between the Dark Matter candidate and the SM are fully controlled by the $y_{\mathrm{DM}}$ coupling.
Although this scenario can be phenomenologically similar to the minimal supersymmetric standard model (MSSM) with a Bino LSP and a right stop, we point out that in the supersymmetric case the $y_{DM}$ coupling is fixed by the LSP composition and it is of the order of the EW couplings ($y_{DM} \sim 0.1-1$). In the scenario discussed here we assume $y_{DM}$ to be a free parameter, which can be as large as allowed by perturbativity, $y_{DM} \lesssim 4 \pi$.
The Dark Matter implications of this scenario were studied in Ref.~\cite{Garny:2018icg}, where it has been shown that the correct Dark Matter relic density can be achieved for a wide range of mass values: $10 \mbox{ GeV} \lesssim m_\chi < 50 \mbox{ TeV}$ and $\Delta M \lesssim 500 \mbox{ GeV}$ as long as  the value of $y_{DM}$ is properly chosen. In particular, large coupling values, e.g. $y_{DM} \gtrsim 3$, are needed in some regions of parameter space. In this work we do not impose any Dark Matter constraints, since these can be modified by the presence of additional (heavy) BSM states and/or a non-standard cosmological evolution.

The above model can lead to several implications at the LHC and low energy observables.
In this work we are mostly interested in the complementarity between direct searches for the top scalar and constraints from top pair production observables.  
A full study of the direct and indirect constraints on the BSM model is left for a future work.

\subsubsection{EFT Limit}
\label{sec:eftmatching}

In the heavy mass limit ($m_T,m_\chi \gg m_t,\sqrt{s}$) the BSM contributions for the model defined in Sec.~\ref{sec:uvmodel} can be described by an effective field theory, where the colored scalar and dark fermion have been integrated out. In this case we have the following dimension-six effective Lagrangian:
\begin{eqnarray}
    \mathcal{L}_{EFT} = m_t C_{g}\; G_{\mu\nu}^A \left( \bar{t} T^A \sigma^{\mu\nu} t \right) & + & C_{q} \left(\bar{t}_R T^A \gamma^\mu t_R \right) \left(\bar{Q}_L T^A \gamma^\mu Q_L +  \bar{u}_R T^A \gamma^\mu u_R + \bar{d}_R T^A \gamma^\mu d_R\right)\nonumber \\
    & + & C_{q} \left(\bar{t}_R T^A \gamma^\mu t_R \right) \left(\bar{Q}_{3,L} T^A \gamma^\mu Q_{3,L}\right) \nonumber\\
    & + & C_{tR} \left(\bar{t}_R T^A \gamma^\mu t_R \right) \left(\bar{t}_R T^A \gamma^\mu t_R \right)\label{eq:lEFTonshell}
\end{eqnarray}
where $u,d,Q$ represent any light quark flavor, $Q_{3,L}$ represents the 3rd generation quark doublet and $m_t$ is the (on-shell) top mass.

The connection with the SMEFT operators in the Warsaw basis described in Sec.~\ref{sec:topeft} and the two operators $C_{g,q}$ is as follows
\begin{eqnarray}\label{ec:smefttoC}
    C_g  &=& y_t^{-1} \frac{C_{tG}}{\Lambda^2},  \nonumber \\ 
    C_q  &=& \frac{C_{tq}^{(8)}}{\Lambda^2}=\frac{C_{t(u/d)}^{(8)}}{\Lambda^2} \nonumber \\
    C_{tR}  &=& \frac{C_{tt}^{(8)}}{\Lambda^2} \nonumber , 
\end{eqnarray}
where $y_t$ is the top Yukawa.
Unlike the general SMEFT framework, the $C_g$ and $C_q$ coefficients are correlated and determined by the underlying UV parameters. These coefficients were computed using Matchete~\cite{Fuentes-Martin:2022jrf} and are given by:
\begin{eqnarray}
    C_g & = &  - \frac{g_s y_{DM}^2}{384 \pi^2} \frac{1}{m_T^2} \frac{1}{\left( 1-x \right)^4}\left[ 1 - 6 x + 3 x^2 + 2 x^3 - 6 x^2 \log(x) \right] \label{eq:Cg}\\
    C_q & = &  \frac{g_s^2 y_{DM}^2}{576 \pi^2} \frac{1}{m_T^2} \frac{1}{\left( 1-x \right)^4}\left[ 2 - 9 x + 18 x^2  - 11 x^3 + 6 x^3 \log(x) \right] \label{eq:Cq} \\
    C_{tR} & = &  -\frac{y_{DM}^4}{128 \pi^2} \frac{1}{m_T^2} \frac{1}{\left( 1-x \right)^3}\left[ 1 - x^2 +2 x \log(x) \right] \label{eq:CtR}
\end{eqnarray}
where $m_t$ is the top mass and $x \equiv m_\chi^2/m_T^2$. 
For example, taking the limit $x\to 1$ ($m_T \simeq m_\chi$), one finds that this model produces a particular pattern in the SMEFT parameter space:
\begin{align}
    C_g \simeq &  - \frac{1}{2}\frac{g_s y_{DM}^2}{384 \pi^2} \frac{1}{m_T^2} \mbox{ , } C_q \simeq \frac{3}{2} \frac{g_s^2 y_{DM}^2}{576 \pi^2} \frac{1}{m_T^2} \mbox{ , } C_{tR} \simeq -\frac{1}{3} \frac{y_{DM}^4}{128 \pi^2} \frac{1}{m_T^2} \label{eq:coeffsDegenerate}\\
    \Rightarrow C_q = & -2 \, g_s \, C_g \label{eq:coeffRatio}
\end{align}

We point out that for the results discussed 
later we only consider BSM contributions up to order $y_{DM}^2$ and the $C_{tR}$ operator will be ignored. Furthermore, the 4-top operator $C_{tR}$ contribution to the top pair production at the LHC is negligible, as it corresponds to a two-loop contribution, and its inclusion would not change the results presented in this work. Nonetheless, this four-top operator could be searched for in the four-top final state, which has recently been observed at the LHC~\cite{CMS:2023ftu,ATLAS:2023ajo}. 
Despite the strong $y_{DM}^4$ scaling, the bounds on our benchmark would be weaker than those from $t\bar t$ final state, due to the kinematic suppression due to a top radiating three tops from $C_{tR}$ and the inherent loop suppression.  Note, also, that a large four-top operator is constrained by perturbativity and, in particular, by the absence of $t \bar t$ bound states. As it was first discussed  in Ref.~\cite{Bardeen:1992mhn}, bound states would be formed when four-fermion interactions like $C_{tR}$ grow above some critical value, estimated to be $C_{tR} \simeq  \frac{8 \pi^2}{3 \Lambda^2}$, where $\Lambda$ represents the scale at which confinement occurs. Comparing  with the matching in Eq.~\ref{eq:coeffsDegenerate} and approximating $\Lambda\simeq m_T$, non perturbativity would require $y_{DM}\gtrsim 18$. Therefore, for the results presented below, we impose $y_{DM} \leq 10$.

\begin{figure}[!h]
    \centering
    \includegraphics[scale=0.49]{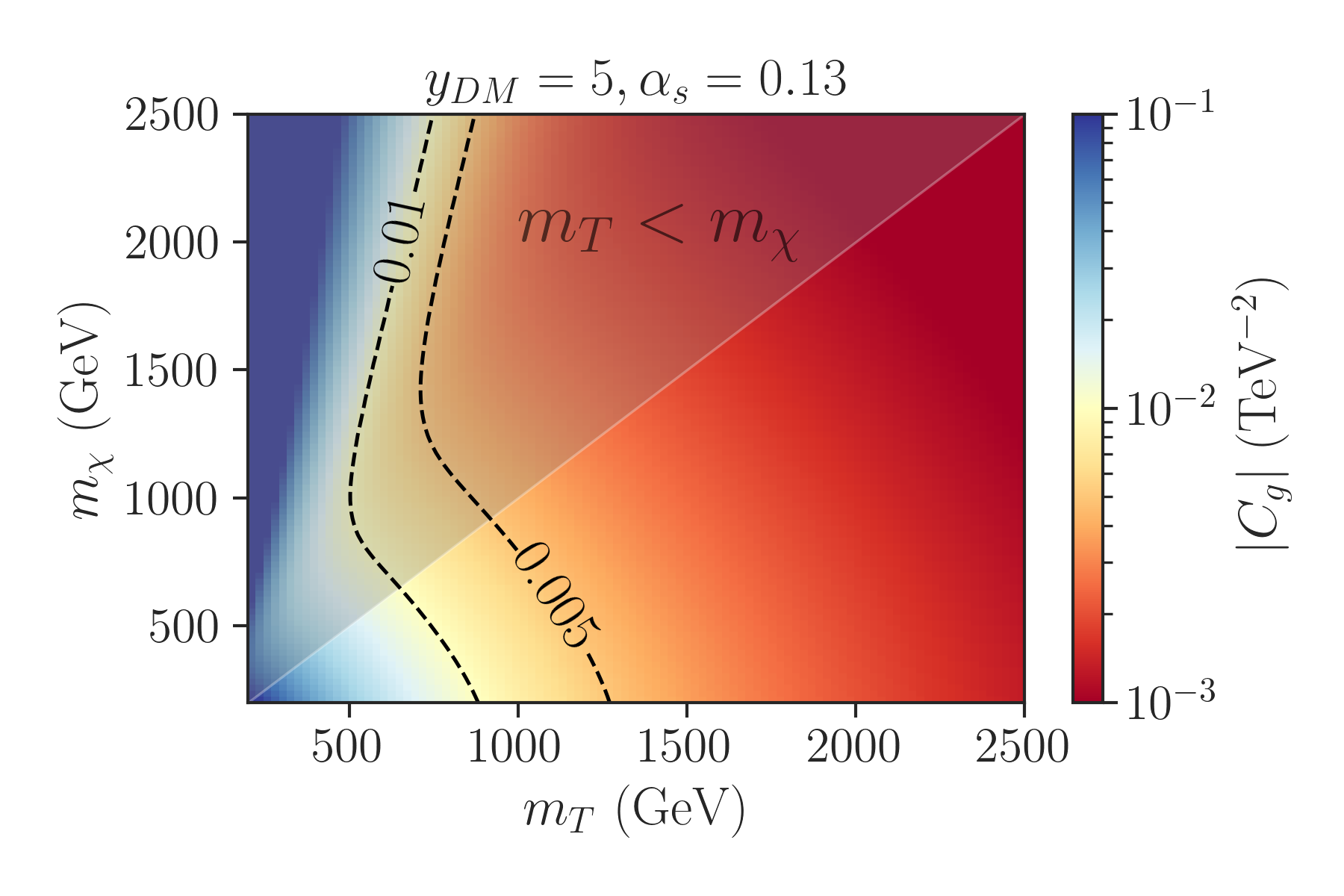}
    \includegraphics[scale=0.49]{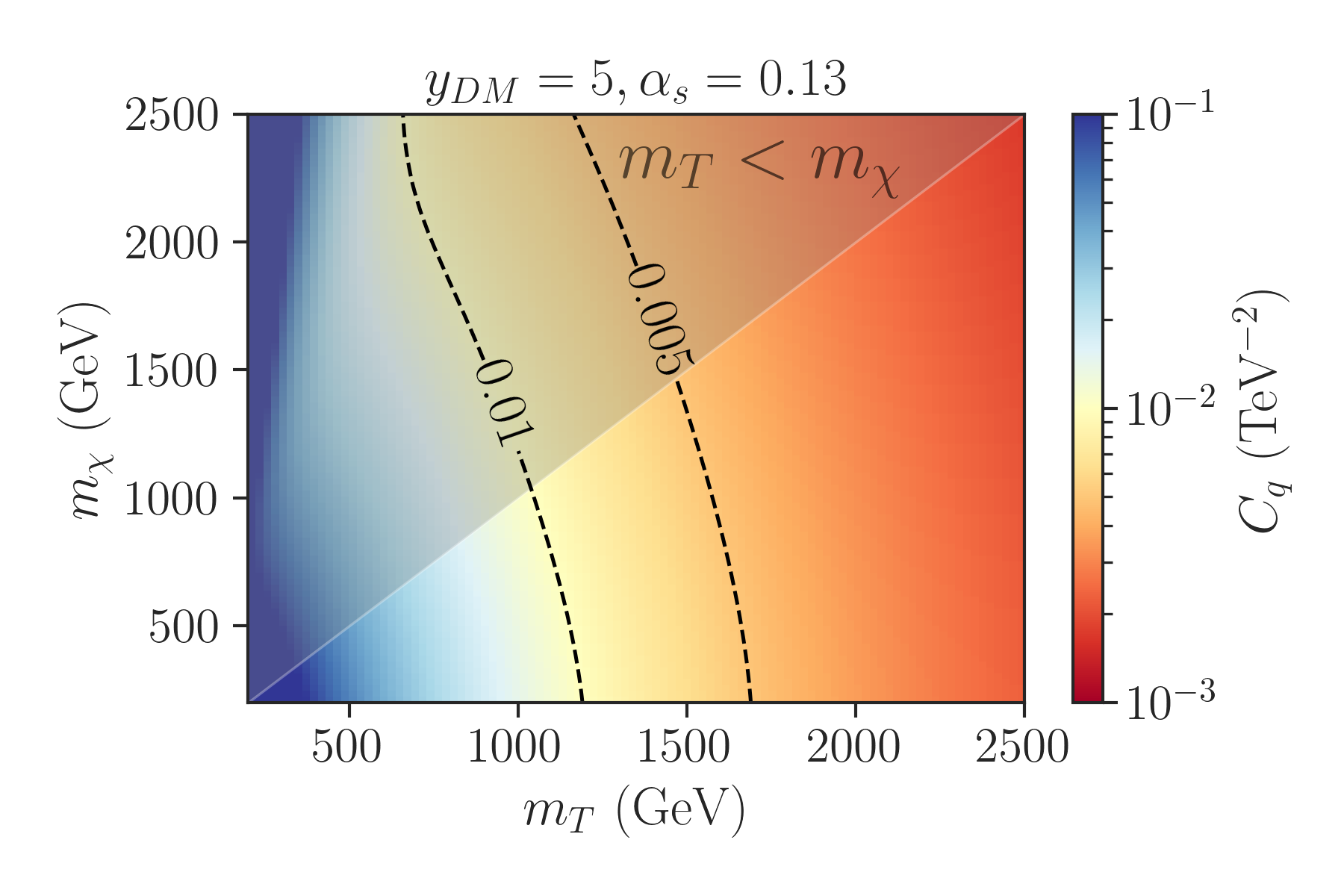}
    \includegraphics[scale=0.49]{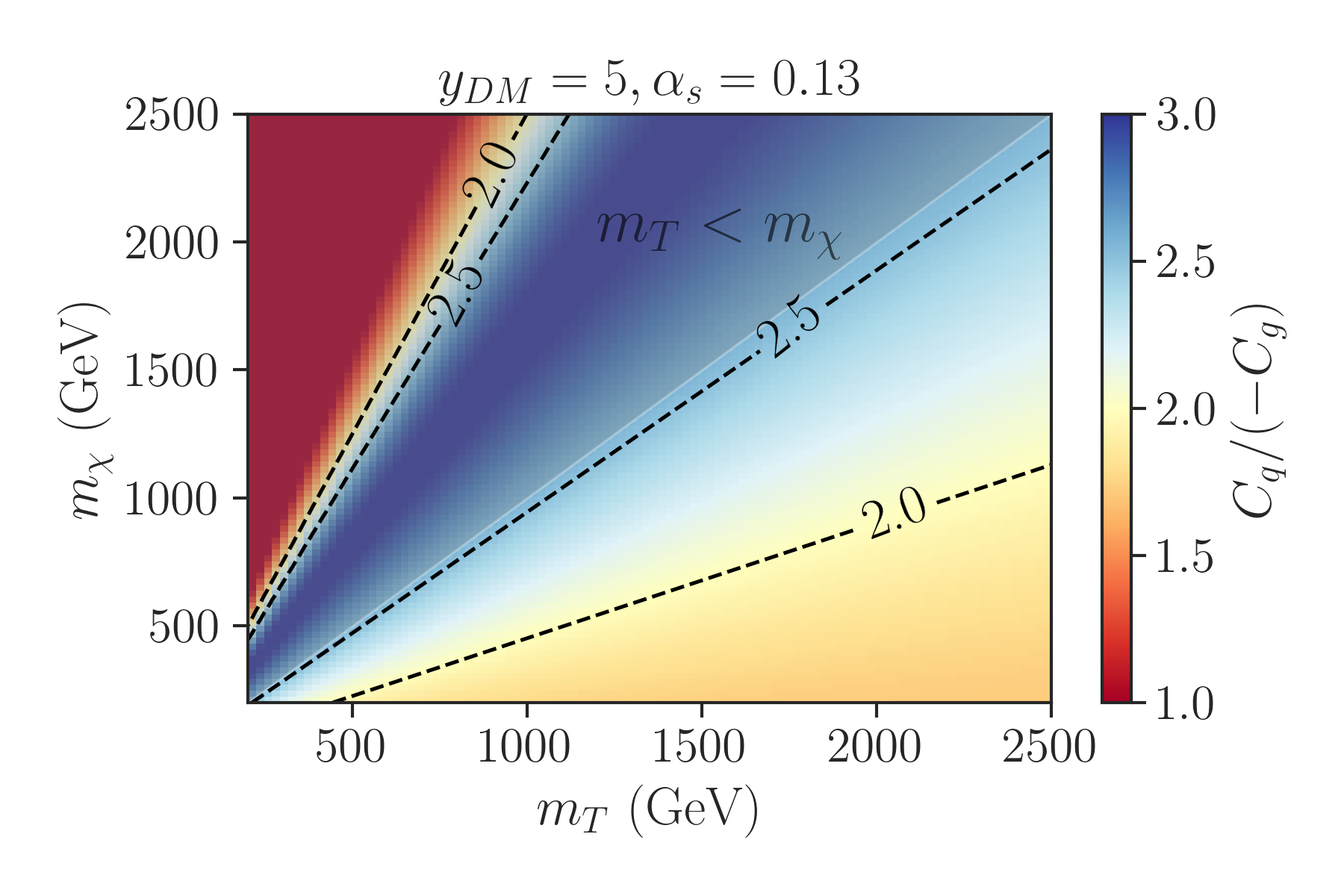}
    \caption{Values for the EFT coefficients $C_g$ and $C_q$ computed according to Eqs.(~\ref{eq:Cg}) and (~\ref{eq:Cq}). The BSM and strong couplings were taken as $y_{DM} = 5$ and $\alpha_s = 0.13$. The shaded region corresponds to $m_T < m_\chi$, which corresponds to a stable colored scalar. The bottom plot shows the values taken by the ratio of both coefficients.}
    \label{fig:coeffs}
\end{figure}

The size of the $C_q$ and $C_g$ coefficients  and their ratio is illustrated in Fig.~\ref{fig:coeffs}. As we can see, for BSM masses around 1 TeV and $y_{DM} = 5$, the coefficients are $C_{g,q} \sim 10^{-2}$~TeV$^{-2}$. Furthermore, we see that $C_q$ is typically $\sim 2-2.5$ times larger than $|C_g|$ and $C_g$ is always negative. As discussed in Sec.~\ref{sec:topeft}, usual SMEFT analysis constrain these coefficients to $0.1-1$~TeV$^{-2}$. Although these constraints can not be directly applied to our scenario, since both coefficients are present and correlated, one would still expect that, for $y_{DM} \lesssim 5$, only the sub-TeV region of parameter space can be tested.

\subsubsection{1-Loop Form Factors}\label{sec:formFactors}

The EFT approach discussed in Sec.~\ref{sec:eftmatching} is only valid for energies well below the BSM masses. As we will show in Sec.~\ref{sec:lhc}, at the LHC it is possible to probe distributions at energies up to a few TeV. Therefore the EFT validity is not guaranteed when using such measurements to look for new physics.
In this case we need to compute the full loop contributions to the relevant observables, which are valid at any scale. For sufficiently high values of BSM masses, the loop contributions should reproduce the EFT results.

\begin{figure}
    \centering
    \includegraphics[scale=1]{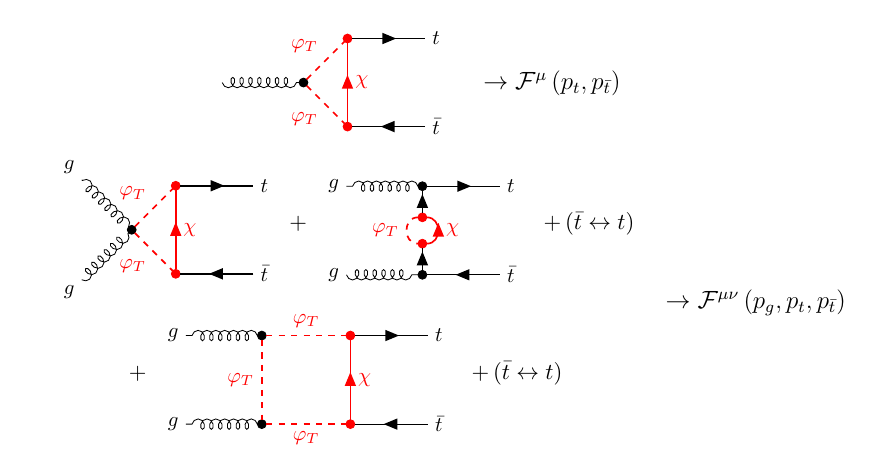}
       \caption{Loop diagrams used to compute the form factors defined in Eq.(~\ref{eq:formFactors}).}
    \label{fig:formFactors}
\end{figure}

In order to compute the 1-loop contributions to top pair production distributions, we have computed form factors for the effective top-top-gluon and top-top-gluon-gluon couplings induced by the loop diagrams shown in Fig.~\ref{fig:formFactors}. The form factors can be then written as effective, momentum dependent couplings:
\begin{equation}
    \mathcal{L}_{FF} = \pi^2 g_s y_{DM}^2 G_\mu \bar{t}\left[ \mathcal{F}^{\mu}\left(p_t,p_{\bar{t}}\right)\right] t + \pi^2 g_s^2 y_{DM}^2 G_\mu G_\nu \bar{t}\left[ \mathcal{F}^{\mu\nu}\left(p_g,p_t,p_{\bar{t}}\right)\right] t 
 \label{eq:formFactors}
\end{equation}
where the $\mathcal{F}^\mu$ and $\mathcal{F}^{\mu\nu}$ form factors contain the full momenta dependence as well as the Dirac and color structures, which we suppress for simplicity.
In order to determine these functions, all the diagrams shown in Fig.~\ref{fig:formFactors} were computed using FeynArts~\cite{Hahn:2000kx} and FeynCalc~\cite{Shtabovenko:2020gxv,Shtabovenko:2016sxi,Mertig:1990an} 
and the results were then used to extract $\mathcal{F}^\mu$ and $\mathcal{F}^{\mu\nu}$. The form factors also include the counter-terms required for renormalizing the top self-energy and the top-top-gluon vertex, which were computed using NLOCT~\cite{Degrande:2014vpa} under the on-shell renormalization scheme.

While $\mathcal{F}^{\mu\nu}$ contains quite a large number of terms and an involved tensor structure, the expression for the $g-t-\bar{t}$ form factor can be written in a compact form using the triangular Passarino-Veltmann loop functions~\cite{Passarino:1978jh,tHooft:1978jhc}:
\begin{align}
    \mathcal{F}^\mu = i T^A & \biggl\{ \slashed{p}_1 \left[ p_1^\mu \left( C_1 + 2 C_{11} \right) - p_2^\mu \left( C_1 + 2 C_{12} \right) \right]  \nonumber \\
    &  + \slashed{p}_2 \left[ p_2^\mu \left( C_2 + 2 C_{22} \right) - p_1^\mu \left( C_2 + 2 C_{12} \right) \right]  \nonumber \\ 
    & + 2  \gamma^\mu \left(C_{00} + \delta_{R} \right) \vphantom{ \left[ \left( p_1 \right) \right]} \biggr\}  P_R + 2 i T^A \gamma^\mu \delta_{L} P_L  \label{eq:fmu}
\end{align}
where $P_{R,L} = \frac{1}{2}\left(1 \pm \gamma^5 \right)$, $p_1, p_2$ are the top and anti-top momenta and $s = (p_1+p_2)^2$. All the loop integrals $C_i,C_{ij}$ are functions of $(p_1^2,s,p_2^2)$ and $\delta_{L,R}$ are the counter-terms obtained using the on-shell renormalization scheme (see Appendix~\ref{app:formFactors} for more details). Although the analytical expressions for the loop integrals are quite involved in the general case, in the mass degenerate limit ($m_\chi \simeq m_T$) and neglecting the top mass ($m_t \ll m_T,m_\chi,\sqrt{s}$) they simplify considerably, resulting in:
\begin{align}
    \mathcal{F}^\mu = &i \frac{T^A}{32 \pi^4}  \left[ 2\left(\slashed{p}_2 p_1^\mu + \slashed{p}_1 p_2^\mu \right) - s \gamma^\mu \right]  P_R \mathcal{F}(s,m_T^2)  \; \mbox{ ($m_\chi \to m_T$, $m_t \to 0$)} , \label{eq:fmuZero}
\end{align}
where:
\begin{align}
    \mathcal{F}(s,m_T^2) \equiv & \frac{1}{s^2} \left[ m_T^2 L^2(s,m_T^2) + \sqrt{s\left(s - 4 m_T^2\right)}  L(s,m_T^2)  + 3 s\right] \mbox{ and } \label{eq:fmuZero2} \\
    L(s,m_T^2) \equiv &  \log\left[\left(\sqrt{s\left(s - 4 m_T^2\right)} + 2 m_T^2 -s \right)/(2 m_T^2) \right] \nonumber
\end{align}

\begin{figure}[!t]
    \centering
    \includegraphics[scale=0.45]{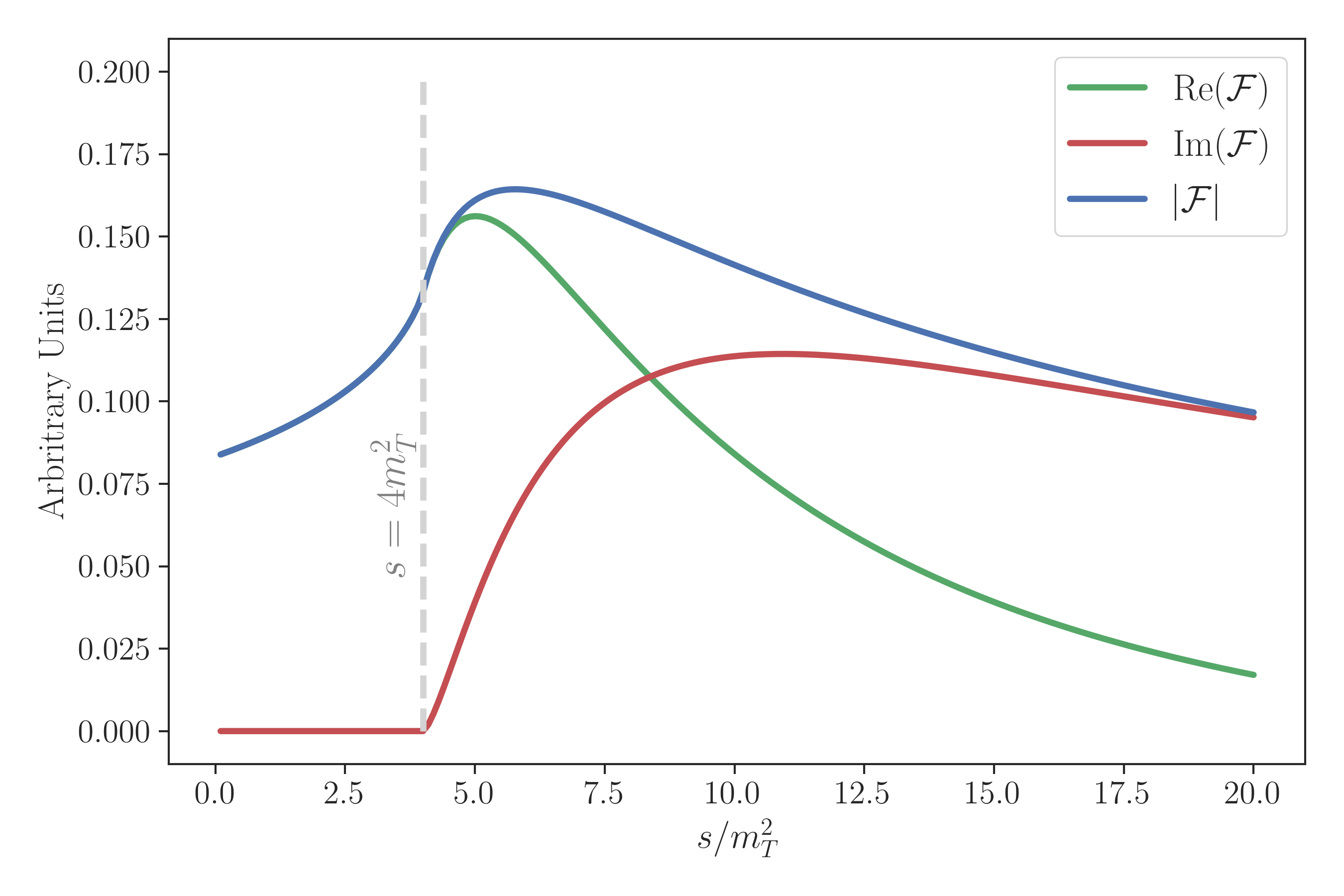}
    \caption{Dependence of form factor on the top pair invariant mass $s  = m(t\bar{t})^2$ for $m_\chi = m_T$ and $m_t \to 0$.
    The curves correspond to the real, imaginary and absolute values of the factor $\mathcal{F}$ defined in Eq.(\ref{eq:fmuZero2}).}
    \label{fig:formFactorLimit}
\end{figure}

In Fig.~\ref{fig:formFactorLimit} we show the form factor dependence on $s$ coming from the $\mathcal{F}(s,m_T^2)$ factor defined in  Eq.(~\ref{eq:fmuZero2}). As we can see, $\mathcal{F}$ peaks around $s \simeq 5 m_T^2$ or $m(t\bar{t}) \simeq \sqrt{5} m_T$, displaying a very broad resonant behavior. We also see that for $s \gg m_T^2$ the form factor is dominated by its imaginary part, resulting in a negative interference term, as discussed below.

\subsection{Differential distributions in $t \bar t$}\label{sec:differential}

Since the BSM model discussed here mostly couples to the top quark, it can impact top pair production and its measured distributions.
In order to illustrate these effects, we consider the top pair invariant mass $m(t\bar{t})$ and the top transverse momentum $p_T(t)$. 
We have implemented the lagrangians defined in Eqs.(\ref{eq:lEFTonshell}) and (\ref{eq:formFactors}) in the UFO~\cite{Degrande:2011ua} format, which allows us to generate events using either the full 1-loop calculation or the EFT approximation. 
For all the results discussed below we have produced 150k MC events for $p p \to t \bar{t}$ at parton level using {\tt MadGraph5\_aMC@NLO}~\cite{Alwall:2014hca,Frederix:2018nkq} with an invariant mass bias, so we can appropriately describe the high energy tail of the distributions. The top quarks were then decayed using MadSpin~\cite{MadSpin}. We have used the PDF set {\tt NNPDF23\_nlo\_as\_0119}  and the factorization and normalization scales were set to the top transverse mass: $\mu_F = \mu_R = \sqrt{m_t^2 + p_T^2}$.

The distributions were computed at leading order in $\alpha_s$ and $y_{DM}$ ($\mathcal{O}(\alpha_s,\alpha_s y_{DM}^2)$), which corresponds to the Born plus the interference terms:
\begin{equation}
    |\mathcal{M}|^2 = |\mathcal{M}_{\rm SM}|^2 + 2 \mathrm{Re}\left( {\mathcal{M}^*_{\rm SM}  \mathcal{M}_{\rm BSM}} \right) \label{eq:amp}
\end{equation}
In the EFT approach this is equivalent to keeping only the  $1/\Lambda^2$ terms.\footnote{We have verified that the contribution from the $1/\Lambda^4$ term ($|\mathcal{M}_{\rm BSM}|^2$) is always subdominant for perturbative values of the BSM coupling, $y_{DM} \lesssim 4 \pi$.} We also point out that the interference term can be negative or positive depending on the behavior of the form factors at distinct energy scales. Since the quark initiated process ($q \bar{q} \to t \bar{t}$) is only affected by $\mathcal{F}^{\mu}$, while the gluon process ($g g \to t \bar{t}$) depends on  both form factors, it is interesting to investigate the individual contributions from each process.
For instance, in the EFT regime, we have $C_g < 0$, resulting in a negative interference contribution from $g g \to t \bar{t}$.

\begin{figure}[!t]
    \centering
    \includegraphics[width=0.48\textwidth]{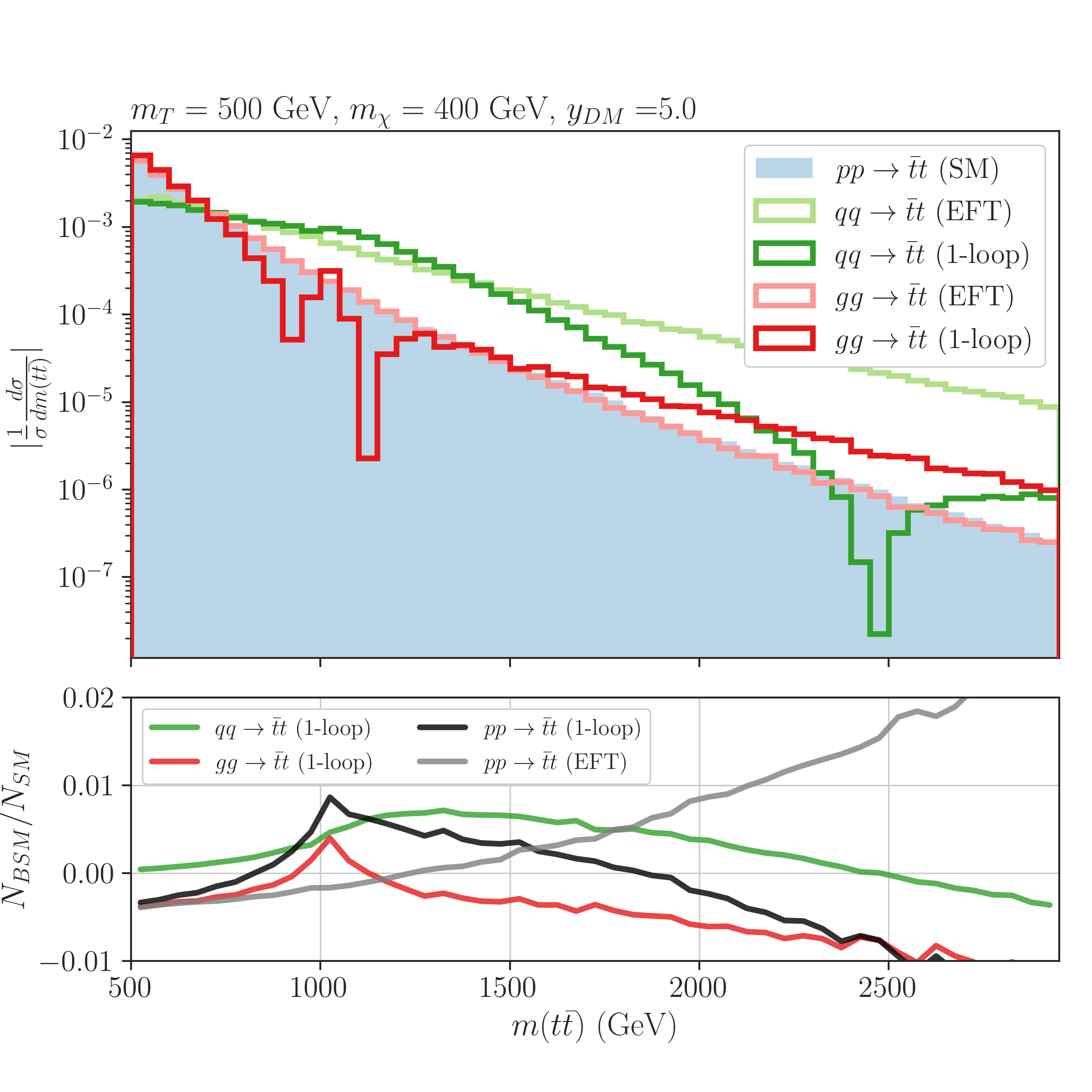}
    \includegraphics[width=0.48\textwidth]{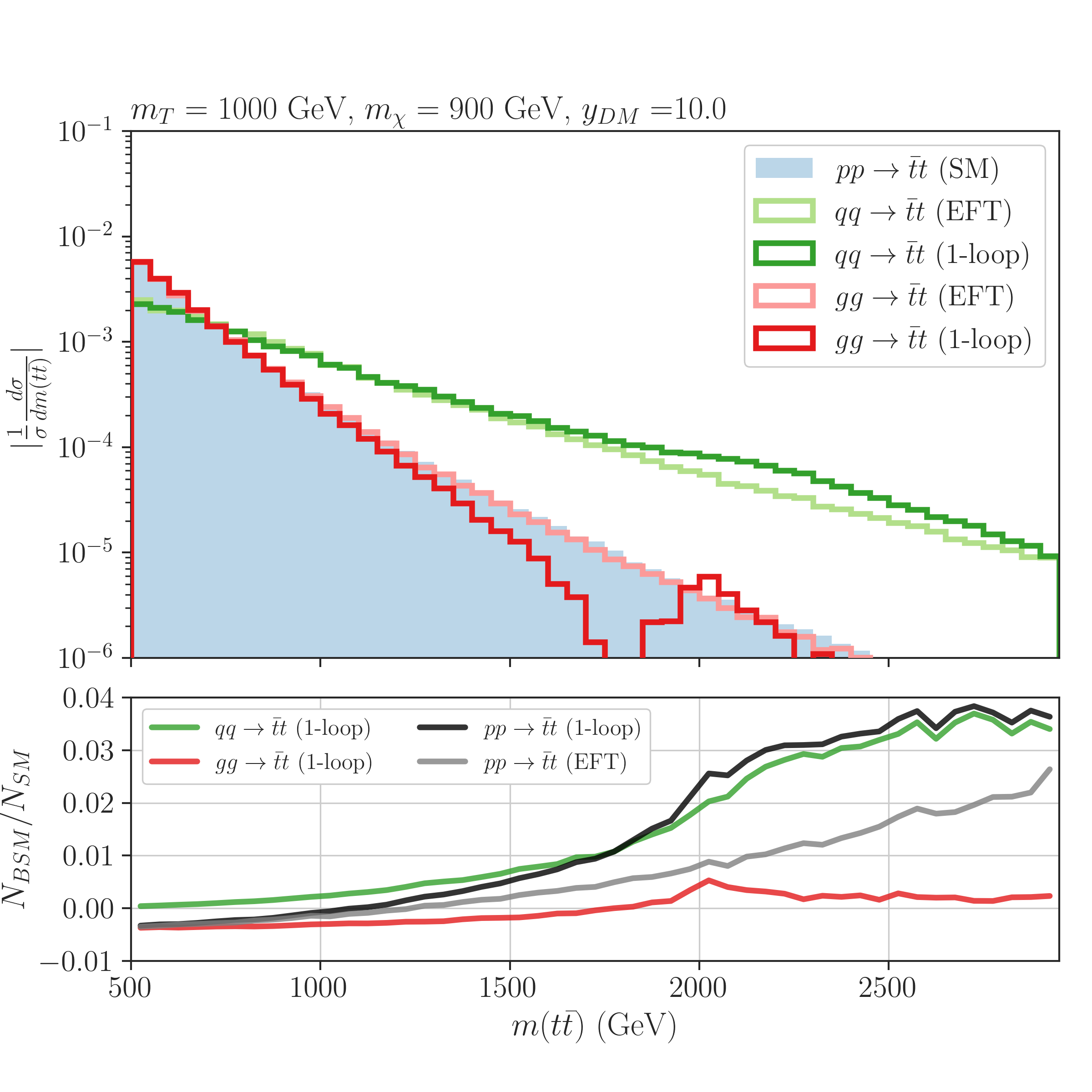}
    \caption{Distributions for the $t \bar{t}$ invariant mass. The SM (Born) contribution is shown by the filled histogram, while the solid histograms show the BSM contributions (interference term) using the full 1-loop calculation and the EFT approximation. The left plot shows the distributions for low BSM masses ($m_T = 500$ GeV and $m_{\chi} = 400$ GeV) and the right plot for higher masses ($m_T = 1000$ GeV and $m_{\chi} = 900$ GeV). The bottom subplots show the ratio of the expected number of events for the BSM contributions to the SM one.}
    \label{fig:mttdists}
\end{figure}

In Fig.~\ref{fig:mttdists} we show the (normalized) $m(t\bar{t})$ distributions for two sets of BSM masses and $y_{DM}$.
The filled histogram shows the SM LO distribution, while the BSM interference term from Eq.(~\ref{eq:amp}) is shown by the solid histograms. The dark red and green histograms correspond to gluon and quark initiated processes computed using the full 1-loop form factors, while the light red and green ones show the results using the EFT approximation. Since the interference can be negative, the upper plots show the absolute value of the distributions, while the lower subplots show the ratio of each BSM contribution to the SM result.

For the light mass case (left plot) we see that the EFT and 1-loop curves start to differ around $m(t\bar{t})\sim 600$~GeV. This is expected, since the EFT approximation is clearly not valid for $\sqrt{s} = m(t\bar{t}) \gtrsim m_T,m_\chi$. First we point out that within the EFT approximation the gluon initiated process always follows very closely the SM distribution, thus simply resulting in a rescaling of the total cross-section. For the scenario investigated here, this contribution is always negative, since $C_g < 0$. The behavior of the 1-loop distribution is also negative for most values of the invariant mass, except for $m(t\bar{t}) \sim 2 m_T$, where the distribution resembles a broad resonance and its contribution becomes positive.
The contribution from the quark initiated process ($q q \to t \bar{t}$), on the other hand, is always  harder than the SM one for both the EFT and 1-loop distributions. Within the EFT approximation, however, the BSM contribution always increases with $m(t\bar{t})$ (relative to the SM), while the 1-loop distribution presents a very broad enhancement around $m(t\bar{t}) = 2 m_T = 1$~TeV. 
In addition, for large invariant mass values ($m(t\bar{t}) \gtrsim 2.5$~TeV), the 1-loop $q q \to t \bar{t}$ contribution becomes negative.
These features can be traced back to the discussion on the form factor in the previous section. As seen in Fig.~\ref{fig:formFactorLimit}, $\mathcal{F}$ presents a broad bump behaviour near $\sqrt{5}\, m_T$ and a dominance of the imaginary part for large $s$ values.
As a result, the full 1-loop distribution displays an excess for the invariant mass bins close to $2 m_T$ and the "intermediate" bins would be the most sensitive to BSM contributions.
This behavior would not be expected if we (wrongly) assumed the EFT approximation to hold, since its distributions tend to always grow with $m(t\bar{t})$.
We also see that the EFT approximation considerably overestimates the signal at the tail of the distribution.

Once we consider higher BSM masses, as shown in the right plot of Fig.~\ref{fig:mttdists}, the EFT and 1-loop distributions agree fairly well up to $\sqrt{s} \sim m_T,m_\chi$, as expected.
The broad resonant behavior of the 1-loop distributions is once again present, but now it only starts to appear at $m(t\bar{t}) \simeq \sqrt{5} m_T \simeq 2.2$~TeV.
In this example the higher bins would be the most sensitive to the BSM contributions and the constraints are stronger than the ones expected from the EFT approximation, since the 1-loop distribution is clearly larger than the EFT one at the tail of the distribution. In addition, the gluon and quark initiated processes in the 1-loop calculation are both positive at the tail, while the gluon curve is always negative if we assume the EFT approximation, thus reducing the total BSM EFT signal.

\begin{figure}[!t]
    \centering
    \includegraphics[width=0.48\textwidth]{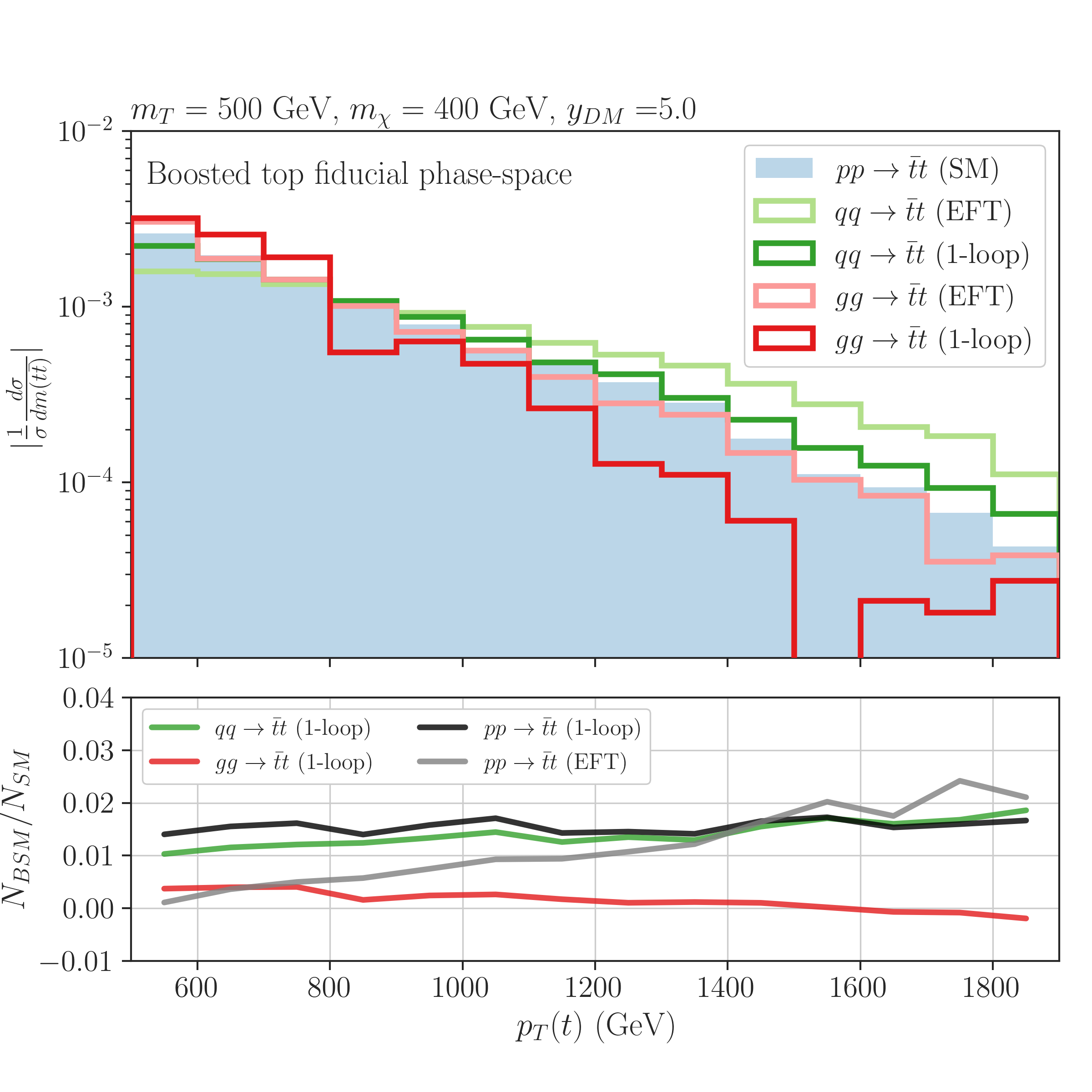}
    \includegraphics[width=0.48\textwidth]{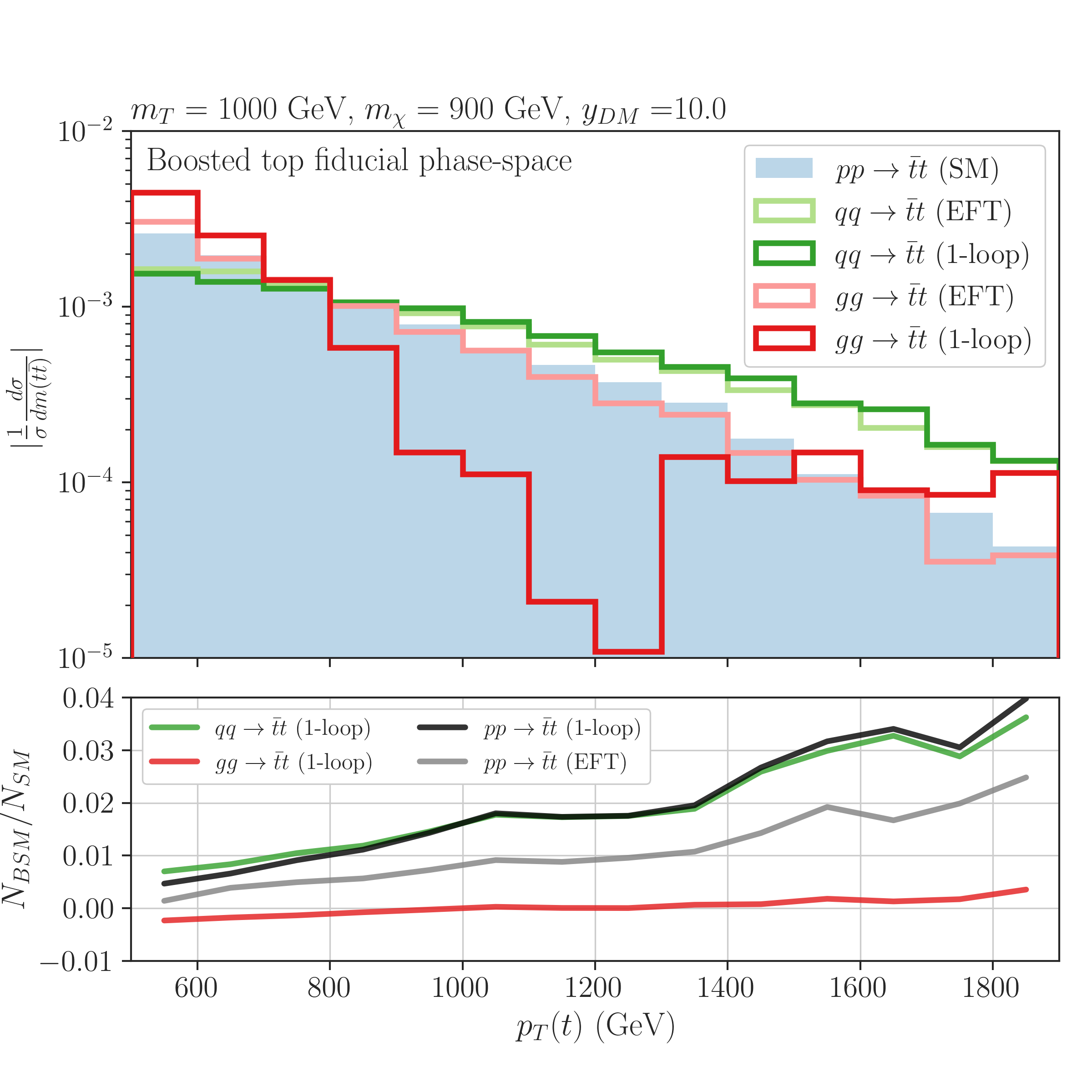}
    \caption{Same as in Figure~\ref{fig:mttdists} for the top transverse momentum. The boosted top phase space was select according to the cuts defined in Table.~\ref{tab:atlasCuts}}
    \label{fig:pTdists}
\end{figure}

In addition to the top pair invariant mass, the top $p_T$ can also be used to constrain new physics contributions~\cite{ATLAS-TOPQ-2019-23}.
Since below we will consider the ATLAS measurement from Ref.~\cite{ATLAS-TOPQ-2019-23} and it includes fiducial phase-space cuts, in Figure~\ref{fig:pTdists} we show the $p_T$ distributions after applying the ATLAS selection. All cuts applied at the particle level are listed in Table~\ref{tab:atlasCuts} and target events where one top decays hadronically and the other leptonically.
The distributions shown in Fig.\ref{fig:pTdists} correspond to the transverse momentum of the hadronically decaying top.
Although the differences between the 1-loop calculation and the EFT approximation are not so dramatic as in the $m(t\bar{t})$ case, we also notice relevant differences between the two methods. In particular, the gluon initiated contributions from the 1-loop results are positive for a wide range of $p_T$ values, while it is always negative within the EFT approximation. 
We also see that total 1-loop result is larger than the EFT approximation up to $p_T \sim 3 m_T$.
It is also important to point out that while the EFT approximation is valid for $\sqrt{\hat{s}} = m(t\bar{t}) \ll m_T,m_{\chi}$, it is not so easy to identify a similar condition for the $p_T$ distribution. This can be seen in Fig.\ref{fig:pTdists} (right), where the EFT approximation fails even at $p_T$ values much smaller than $m_T,m_\chi$.

\begin{table}[h!]
\centering
\begin{tabular}{|l|c|}
\hline
\multicolumn{2}{|c|}{\large Boosted Top Phase-Space}\\ \hline
\multirow{3}{*}{\textbf{Jet Cuts}} & $n(j) > 0$ \\ & $p_T > 36$~GeV \\ & $|\eta| < 2.5$
\\ \hline
\multirow{5}{*}{\textbf{Fat jet Cuts}} & $n(j) > 0$ \\ & $p_T > 355$~GeV \\ & $|\eta| < 2.0$ \\ & 120~GeV $< m < 220$~GeV \\ & contains one $b$
\\ \hline
\multirow{6}{*}{\textbf{Lepton Cuts}} & $n(l) = 1$ \\ & $p_T > 27$~GeV \\ & $|\eta| < 2.5$ \\  & $\Delta R(l,b) < 2.0$ \\ & $m(l,b) < 180$~GeV \\ & $\Delta R(l,j) > 0.4$
\\ \hline
{\textbf{$E_{T}^{\rm miss}$ Cut}} & $E_{T}^{\rm miss} > 20$~GeV\\ \hline
\end{tabular}
\caption{Fiducial phase-space cuts applied at parton level to reproduce the boosted top phase-space considered by ATLAS in Ref.~\cite{ATLAS-TOPQ-2019-23} \label{tab:atlasCuts}.}
\end{table}

\section{LHC Constraints}
\label{sec:lhc}

As shown in Sec.~\ref{sec:differential}, the BSM model considered here can have an impact on the differential top distributions.
However, for masses smaller than a few TeV, the BSM states can be produced on-shell at the LHC. Therefore this scenario can potentially be constrained by: i) direct searches, i.e. searches for on-shell $\phi_T$ production, and ii) indirect searches, i.e. measurements of $p p \to t \bar{t}$ distributions.
In this Section we mostly aim to address the following questions:
\begin{itemize}
    \item Can indirect searches be complementary to direct searches?
    \item What is the impact of (wrongly) assuming the EFT approximation when constraining the model?
\end{itemize}
Clearly the above answers depend on the model parameters: $(m_T,m_\chi,y_{DM})$ or $(m_T,\Delta M = m_T -m_\chi,y_{DM})$.
For instance, for sufficiently large masses we expect the EFT results to be valid. Also, for very small $y_{DM}$, the loop contributions to top pair production are suppressed and direct searches will be more sensitive.

\begin{figure}[!t]
\centering
  \includegraphics[scale=1]{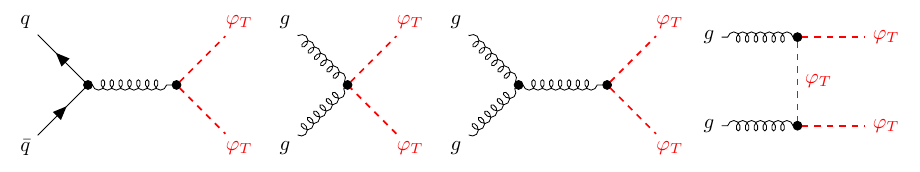}
\caption{Leading order diagrams for the on-shell production of the colored scalar. \label{fig:smsProd}}
\end{figure}

\subsection{Direct Searches} \label{sec:direct}

If $\phi_T$ is sufficiently light ($m_T \lesssim$~TeV), it will be copiously produced at the LHC, since it is colored.
The leading order diagrams for $\phi_T$ production are shown in Figure ~\ref{fig:smsProd}.
The signatures generated by $\phi_T$ production and decay strongly depend on the mass difference $\Delta M = m_T - m_\chi$. For sufficiently large $\Delta M$, the signature corresponds to on-shell tops plus missing energy ($E_T^{\rm miss}$), while in the compressed scenario ($\Delta M \ll m_t$) we have $E_T^{\rm miss}$ plus $b$-jets and additional (soft) leptons and light jets. The compressed scenario tends to be more challenging, resulting in weaker constraints on $m_T$.
We also point out that the signal does not dependent on the BSM coupling $y_{DM}$, except for the scalar width.
Although for sufficiently small widths $y_{DM}$ the scalar can become long-lived, in the following we assume $y_{DM}$ large enough so $\phi_T$ always have prompt decays.

Since the LHC signatures are the same employed on stop searches, 
we make use of {\sc SModelS}~\cite{Kraml:2013mwa,Alguero:2020grj,Alguero:2021dig,MahdiAltakach:2023bdn} 
to reinterpret the ATLAS and CMS constraints on stop-neutralino simplified models and identify the most relevant analyses.
The compressed scenario is particularly challenging and for very small $\Delta M$ the decay products can be very soft and missed by most event selection criteria. In this region of parameter space searches for Dark Matter production, which target initial state radiation (ISR) jets plus $E_T^{\rm miss}$ can become relevant.
Therefore, in addition to the stop searches, we have considered the CMS jets plus $E_T^{\rm miss}$ search~\cite{CMS-EXO-20-004} which targets scenarios with hard jets coming from ISR.
We have recast this analysis and used {\tt MadGraph5\_aMC@NLO}, {\sc Pythia 8.306}~\cite{Pythia8} and Delphes~\cite{Delphes} to reinterpret the CMS results for the BSM scenario from Sec.~\ref{sec:uvmodel}.
Although we have computed the $\phi_T$ production cross-section at leading order (LO), a constant k-factor $k=1.5$ was used to approximate the NNLO+NNLL result~\cite{Beenakker:2016gmf}.

\begin{figure}[!t]
    \centering
    \includegraphics[scale=0.8]{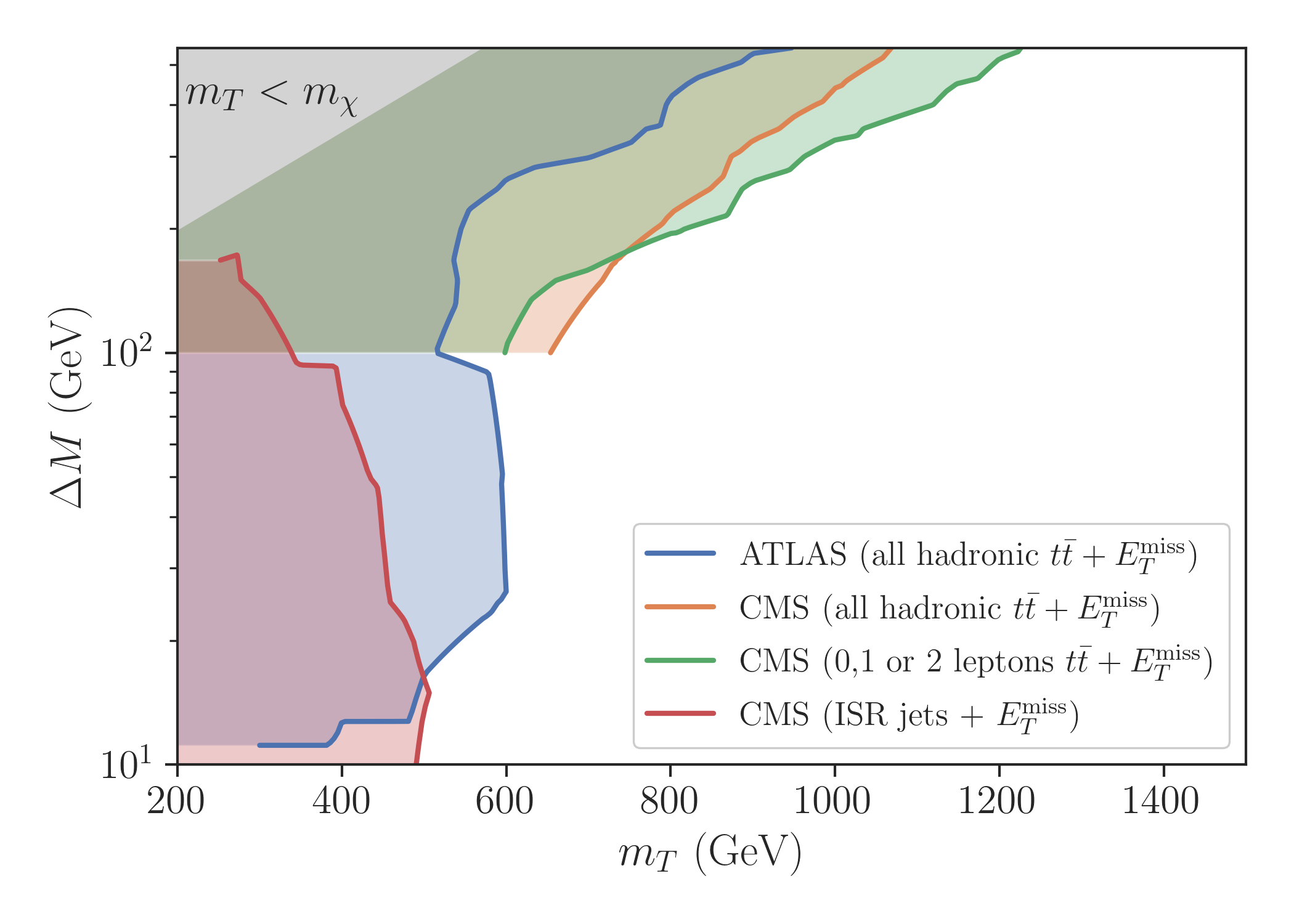}
    \caption{95 \% C.L. exclusions in the $m_T$ versus $\Delta M = m_T - m_\chi$ plane from direct searches for on/off-shell tops and missing energy. The red curve corresponds to the CMS search~\cite{CMS-EXO-20-004} for ISR jets plus $E_T^{\rm miss}$ (EXO-20-004),  the green curve shows the excluded region from the CMS search~\cite{CMS-SUS-20-002} for $0,1$ and 2 leptons plus $E_T^{\rm miss}$ (CMS-SUS-20-002), while the orange and blue curves show the excluded regions from the ATLAS (ATLAS-SUSY-2018-12) and CMS  (CMS-SUS-19-006) searches for hadronic tops plus $E_T^{\rm miss}$ from Refs.~\cite{ATLAS-SUSY-2018-12} and ~\cite{CMS-SUS-19-006}, respectively.}
    \label{fig:exclusionStop}
\end{figure}
In Figure~\ref{fig:exclusionStop} we show the 95\% C.L. excluded region in the $\Delta M$ vs $m_T$ plane.
For large mass differences ($\Delta M > m_W$), the most relevant analyses are the CMS combined stop search~\cite{CMS-SUS-20-002} and the CMS search for jets and missing energy~\cite{CMS-SUS-19-006}. These searches lose sensitivity to scenarios with small $\Delta M$ and limits are not provided in this case, leading to the sharp cut-off seen on the green and orange curves in Fig.~\ref{fig:exclusionStop}.
The ATLAS search for hadronic tops plus missing energy~\cite{ATLAS-SUSY-2018-12}, however, also targets the compressed scenario and is the most sensitive search in this region of parameter space, excluding mass differences down to 10-20 GeV. For even smaller mass differences, the decay products are very soft and the CMS search for ISR jets becomes relevant, as shown by the red curve in Fig.~\ref{fig:exclusionStop}.
Overall we see that in the highly compressed scenario scalar masses up to $m_T \simeq 500$ GeV are excluded, while for large $\Delta M$ the exclusion goes up to 1.3 TeV.

\subsection{Indirect searches in $t \bar t$}\label{sec:indirectSearches}

As discussed in Sec.~\ref{sec:direct}, the constraints from direct searches for $\phi_T$ depend strongly on the scalar-DM mass difference and exclude masses up to $m_T \simeq 1.3$~TeV.
These constraints, however, do not depend on the BSM coupling ($y_{DM}$), except for $y_{DM} \ll 1$, which could render the colored scalar long-lived.
On the other hand, if $y_{DM} \gtrsim 1$, BSM loop contributions to $ p p \to t \bar{t}$ are enhanced and can become sizeable. 
The top distributions have been measured at high accuracy both by CMS and ATLAS and found to be in good agreement with the SM predictions. 
Here we follow closely the approach developed in Ref.~\cite{Esser:2023fdo}, which considered the top pair invariant mass measured by CMS~\cite{CMS-TOP-20-001} and the top transverse momentum measured by ATLAS~\cite{ATLAS-TOPQ-2019-23} to constrain the couplings of axion-like particles to the top quark.

It is well known that the $t\bar{t}$ distributions can be significantly modified by NLO and NNLO QCD corrections~\cite{Catani:2019hip,Catani:2019iny}. Therefore, for the results below, we use the corresponding SM predictions at NNLO quoted by ATLAS or CMS. However, it is beyond the scope of this work to compute the BSM signal to this level of accuracy. Nonetheless, we approximate the impact of higher order corrections on the BSM contribution (interference with the SM) using a bin-dependent reweighting factor:
\begin{equation}
    k_i = \frac{N_{\rm SM}^i ({\rm NNLO})}{N_{\rm SM}^i ({\rm LO})} \Rightarrow N_{\rm BSM}^i ({\rm NNLO}) \simeq k_i N_{\rm BSM}^i ({\rm LO}),
\end{equation}
where $N_{\rm SM}^i ({\rm LO})$ ($N_{\rm BSM}^i ({\rm LO})$) is the background (signal) prediction computed at LO using {\tt MadGraph5\_aMC@NLO}. The reweighting factors obtained through this procedure are typically $k_i \simeq 1.3-1.6$ for the CMS invariant mass bins and $k_i \simeq 0.6-1.8$ for the ATLAS transverse momentum bins.  With the above expressions and the covariance matrices ($C_{ij}$) provided by the experimental collaborations, we can then make use of the measured distributions ($N_{\rm Obs}^i$) to compute limits on the BSM coupling $y_{DM}$. Following the same procedure used by the experimental collaborations we define a $\chi^2$ function as:
\begin{equation}
    \chi^2(y_{DM}) = \sum_{i,j={\rm bins}} \Delta_i C^{-1}_{ij} \Delta_j, 
\end{equation}
where $\Delta_i  = \left[N_{\rm Obs}^i - N_{\rm SM}^i ({\rm NNLO}) - y_{DM}^2 N_{\rm BSM}^i ({\rm NNLO})\right]$, so the 95\% C.L. limit on $y_{DM}$ corresponds to $\Delta \chi^2 = 3.84$.

Below we present results for both the EFT approach discussed in Sec.~\ref{sec:eftmatching} and the full 1-loop form factors from Sec.~\ref{sec:formFactors}.
Although the former should only be valid at high masses, it is interesting to compare both approaches and quantify how the EFT constraints deviate from the full 1-loop calculation.

\subsubsection*{CMS $m(t\bar{t})$}

We first consider the CMS measurement~\cite{CMS-TOP-20-001}  of the differential $m(t \bar{t})$ cross-sections at $\sqrt{s} = 13$~TeV using the full Run 2 luminosity, $\mathcal{L} = 137$ fb$^{-1}$. The measurement includes the full kinematic range and extends up to invariant masses of 3.5~TeV. CMS has unfolded the measured distributions and provided measurements at the parton level, which can be used to constrain BSM contributions. In addition, the covariance matrix of the measurements has been provided, which allows us to include systematic and statistical uncertainty correlations.
For the SM predictions we have considered the NNLO prediction quoted by CMS computed using \textsc{MATRIX}~\cite{Grazzini:2017mhc}, 
but since the covariance matrix for the predictions was not given, we have not included it when computing the limits on the BSM signal.

\begin{figure}[!h]
    \centering
    \includegraphics[width=0.8\textwidth]{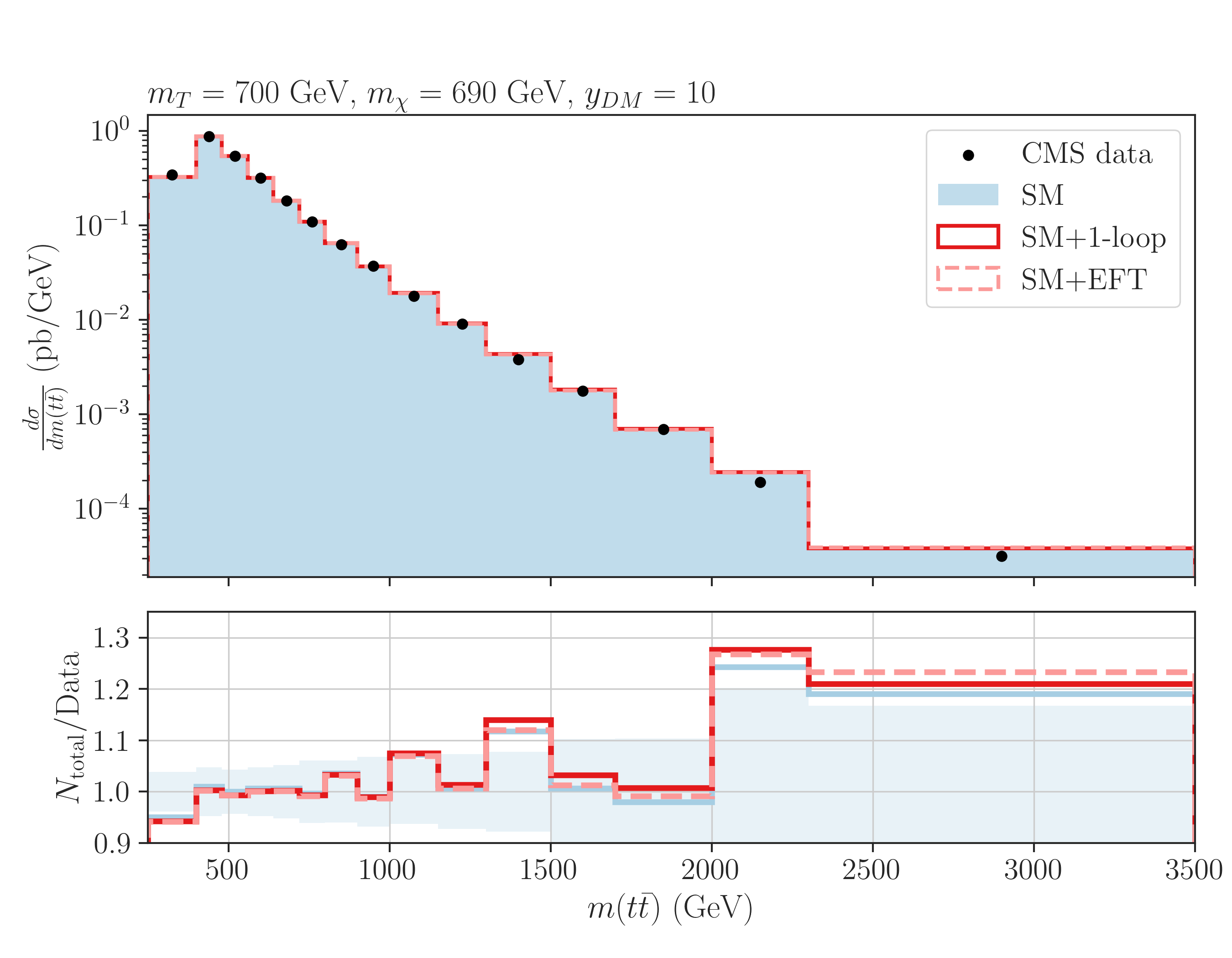}
    \caption{Invariant mass distribution, $m(t\bar{t})$, for $m_T = 700$~GeV, $m_\chi = 690$~GeV and $y_{DM} = 10$. The data points show the unfolded distribution measured by CMS~\cite{CMS-TOP-20-001}, while the filled histogram shows the SM prediction at NNLO~\cite{CMS-TOP-20-001}. The solid histogram shows the SM plus BSM distribution computed using the 1-loop form factors, while the dashed histogram shows the same distribution within the EFT approximation. The bottom subplot shows the ratio of the measured distribution to the SM, SM plus 1-loop and SM plus EFT distributions. The band shows the uncertainties ignoring correlations, i.e. $\sqrt{C_{ii}}$.}
    \label{fig:cmsDists}
\end{figure}

In Fig.~\ref{fig:cmsDists} we show the measured invariant mass distribution, the SM prediction and the SM plus BSM prediction using the 1-loop form factors or the EFT approximation. The bottom subplot shows the ratio of predictions to data. The BSM masses are $m_T = 700$~GeV and $m_\chi = 690$~GeV and correspond to a compressed scenario currently beyond the reach of direct searches.
The BSM coupling is considerably large, $y_{DM} = 10$, but still within the perturbative regime.
As we can see, the 1-loop distribution significantly deviates from data for $m(t\bar{t}) \gtrsim 2 m_T = 1.4$~TeV, as expected from the behavior discussed in Sec.~\ref{sec:differential}. We also note that for the intermediate bins the EFT contribution underestimates the signal, while for the last two bins it is close to the 1-loop calculation. However, since the uncertainty in the last bins is quite high, the BSM signal is mostly constrained by the intermediate bins. As a result, the EFT approximation significantly underestimates the constraints. In particular, for the point shown in Fig.~\ref{fig:cmsDists}, we obtain $y_{DM} < 7.8$ at 95\% C.L. using the 1-loop calculation, while the EFT approximation results in $y_{DM} < 10.7$.
Note that the error band shown in the bottom subplot of Fig.~\ref{fig:cmsDists} corresponds only to $\sqrt{C_{ii}}$. The full correlations, however, are essential for computing the limits and provide stronger constraints than assuming uncorrelated bin uncertainties.

\subsubsection*{ATLAS $p_{T}(t)$}

As discussed in Sec.~\ref{sec:differential}, the top transverse momentum can also be impacted by BSM contributions.
In order to constrain the signal, we consider the ATLAS measurement~\cite{ATLAS-TOPQ-2019-23} of the top $p_T$ for semi-leptonic decaying tops. Unlike the CMS measurement discussed in the previous Section, ATLAS considers the fiducial phase-space for boosted tops, which can be approximated by the cuts listed in Table~\ref{tab:atlasCuts}. The measured distribution is then unfolded to the parton level. For the SM prediction we consider the values quoted by ATLAS obtained using {\tt MadGraph5\_aMC@NLO} and {\sc Pythia 8} after NNLO reweighting~\cite{ATLAS-TOPQ-2019-23}.
The covariance matrix is also provided for the unfolded measurement, but it only includes the statistical uncertainties. However, the total systematical uncertainties for each bin is also given and we include them as a diagonal contribution to $C_{ij}$, which means we ignore correlations of systematical uncertainties\footnote{Although this is clearly an approximation, there is not enough information publicly available to properly include the correlations of the systematical uncertainties.}. 

\begin{figure}[!h]
    \centering
    \includegraphics[width=0.49\textwidth]{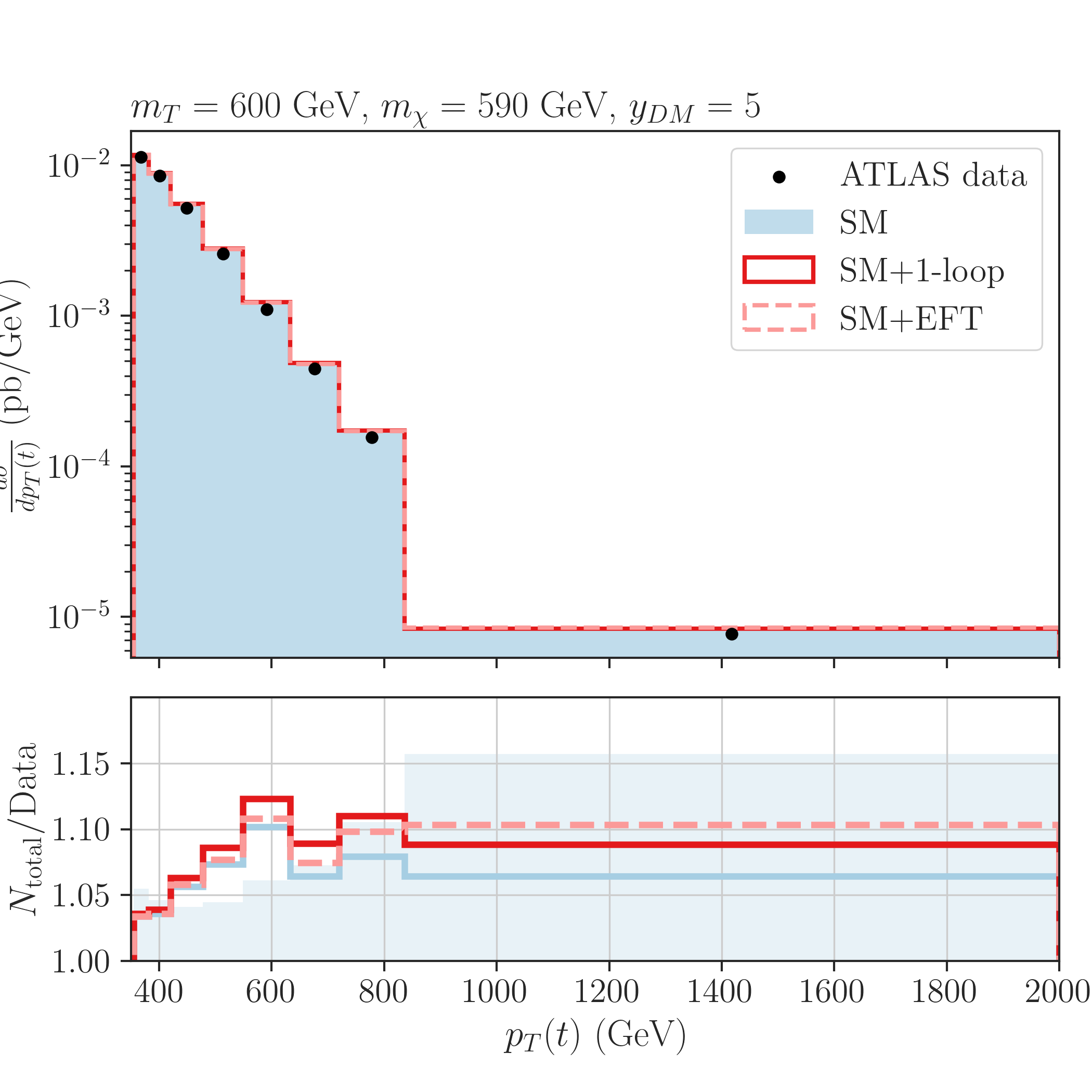}
    \includegraphics[width=0.49\textwidth]{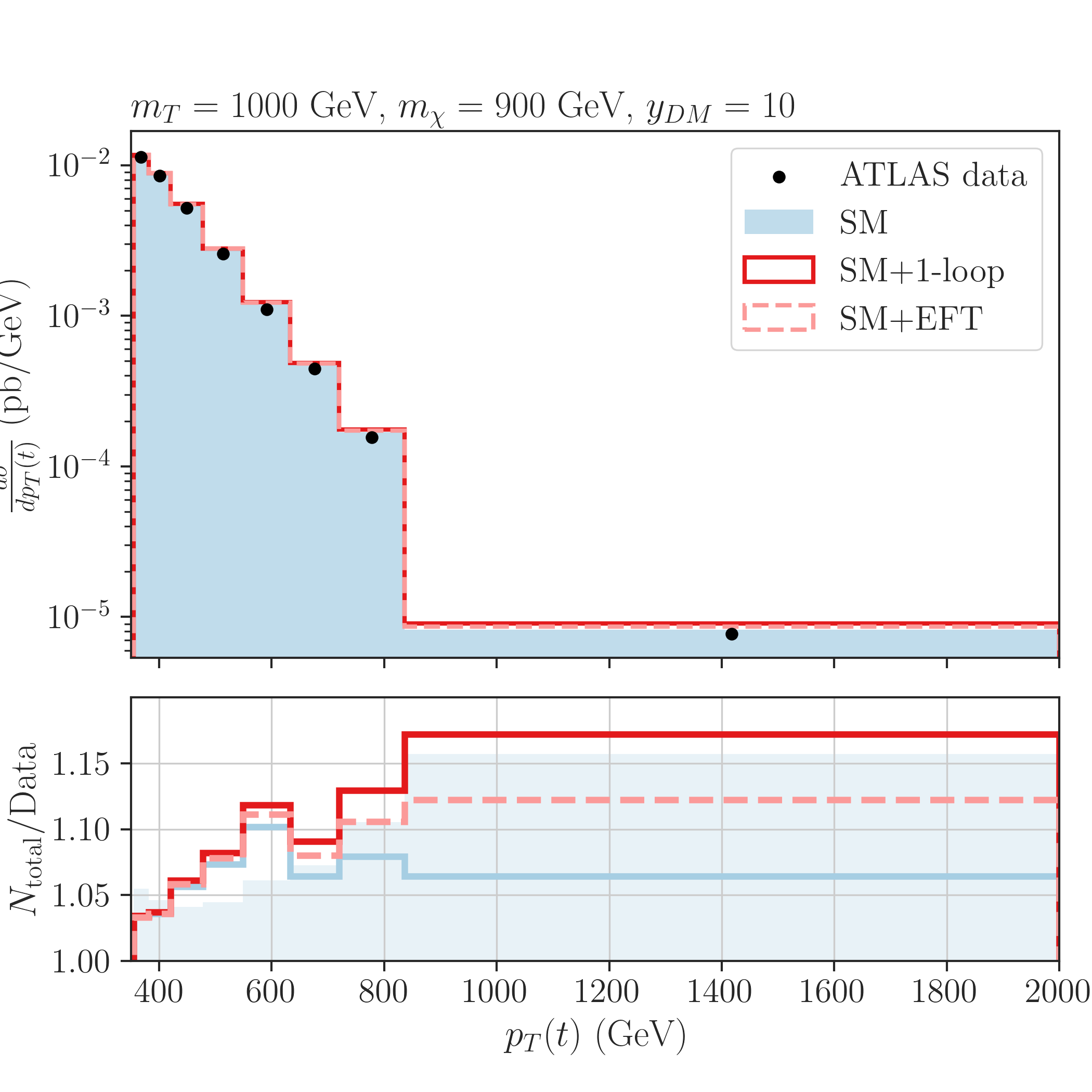}
    \caption{Transverse momentum distribution, $p_T(t)$, for $m_T = 600$~GeV, $m_\chi = 590$~GeV, $y_{DM} = 5$ (left) and $m_T = 1$~TeV, $m_\chi = 0.9$~TeV, $y_{DM} = 10$ (right). The data points show the unfolded distribution measured by ATLAS~\cite{ATLAS-TOPQ-2019-23}, while the filled histogram shows the SM prediction at NNLO from Ref.~\cite{ATLAS-TOPQ-2019-23}. The solid histogram shows the SM plus BSM distribution computed using the 1-loop form factors, while the dashed histogram shows the same distribution within the EFT approximation. The bottom subplot shows the ratio of the measured distribution to the SM, SM plus 1-loop and SM plus EFT distributions. The band shows the uncertainties ignoring correlations, i.e. $\sqrt{C_{ii}}$.}
    \label{fig:atlasDists}
\end{figure}

The measured $p_T$ distribution is shown in Fig.~\ref{fig:atlasDists} along with the SM prediction (filled histogram) and the total distribution (SM plus BSM) computed using the 1-loop form factors and the EFT approximation. The first plot shows the distribution for  $y_{DM} = 5$ and "light" BSM masses, $m_T = 600$~GeV, $m_\chi = 590$~GeV, while the second one shows the same distributions, but for heavier masses ($m_T = 1$~TeV and $m_\chi = 0.9$~TeV) and a larger BSM coupling, $y_{DM} = 10$.
For the lighter BSM masses we see that the 1-loop distribution deviates more strongly from data in the intermediate bins, as expected from the behavior seen in Fig.~\ref{fig:pTdists}. We also see that the EFT underestimates the signal in the intermediate bins, while overestimates it for the highest bin.
Due to the large statistical uncertainties at large $p_T(t)$, the ATLAS measurement is mostly sensitive to deviations in the low to intermediate bins. As a result, the EFT approximation results in weaker constraints to the $y_{DM}$ coupling. In particular, for the signal shown in the left plot of Fig.~\ref{fig:atlasDists}, we obtain $y_{DM} < 4.7$ at 95\% C.L. using the 1-loop calculation, while $y_{DM} < 7.4$ if we assume the EFT approximation.
A similar behavior is also seen at larger masses, as shown by the right plot in Fig.~\ref{fig:atlasDists}. In this case, however, the EFT distribution is smaller than the 1-loop one for all bins and once again underestimates the sensitivity to new physics.

Note that, when compared to the invariant mass distributions from Fig.~\ref{fig:cmsDists}, the $p_T(t)$ measurement seems to be more sensitive to the BSM signal than the invariant mass distribution, since a larger excess is seen in the intermediate $p_T$ bins.
In addition, both measured distributions display under-fluctuations in several bins with respect to the SM prediction. As a result, the constraints on the BSM signal are stronger than expected (see Appendix~\ref{app:expectedLimits} for more details).

\subsection{Results}\label{sec:results}

\begin{figure}[!h]
    \centering
    \includegraphics[width=0.95\textwidth]{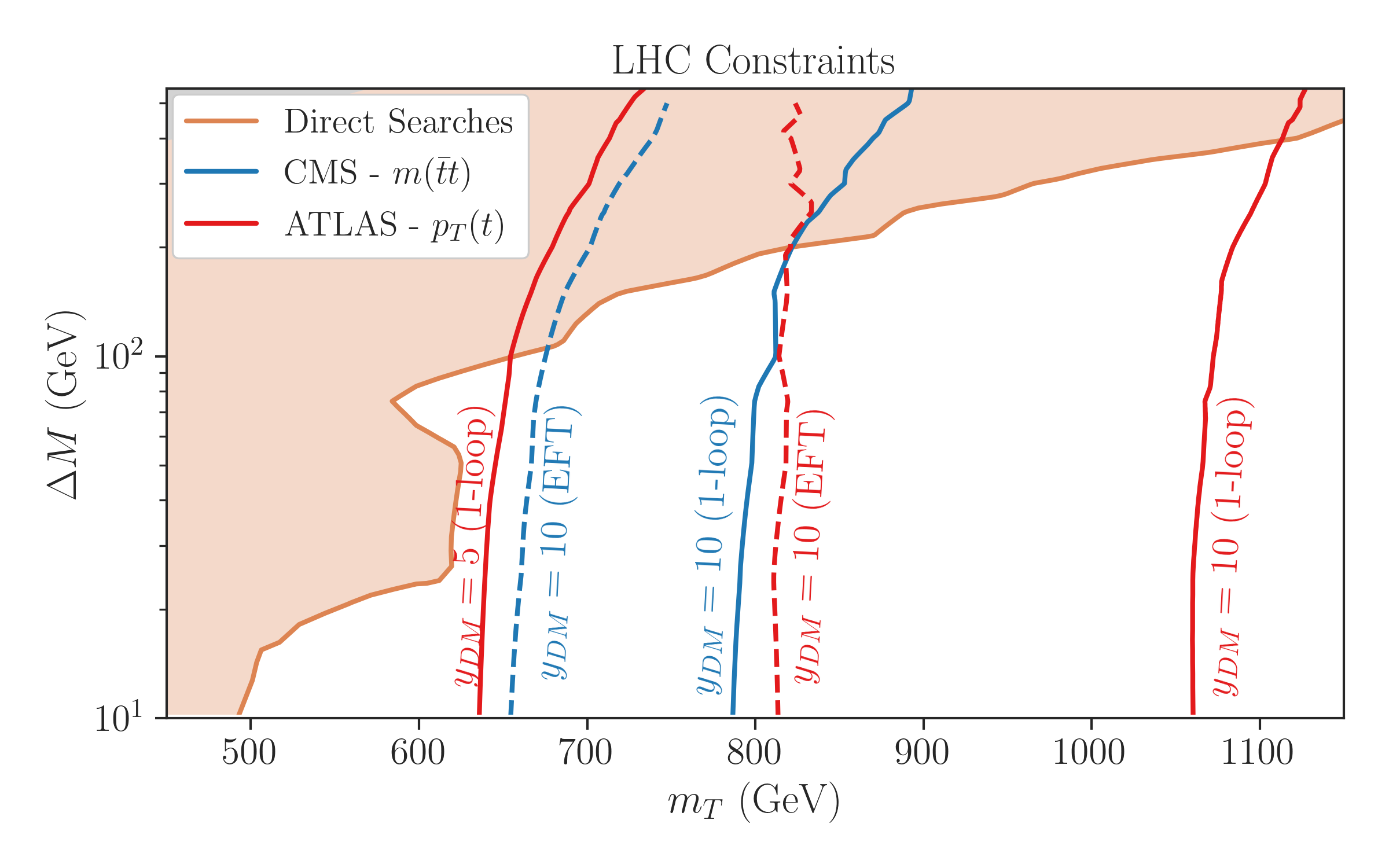}
    \caption{Comparison of the exclusion curves from direct stop searches and from top measurements. The solid lines show the exclusion obtained using the full 1-loop calculation, while the dashed lines correspond to the EFT approximation for distinct values of the BSM coupling $y_{DM}$. The limits obtained using the CMS invariant mass measurement from Ref.~\cite{CMS-TOP-20-001} are shown by the blue curves, while the ones obtained from the ATLAS $p_T$ measurement~\cite{ATLAS-TOPQ-2019-23} are shown in red.}
    \label{fig:allexclusions}
\end{figure}

The discussion in the previous Sections showed that top measurements can be sensitive to new physics and complementary to direct searches, specially in the compressed region, $\Delta M \ll m_t$, and for large BSM couplings, $y_{DM} \gtrsim 5$. In order to compare the constraints from direct and indirect searches, we scan over the BSM masses and compute the limits on $y_{DM}$ obtained from the CMS $m(t\bar{t})$ measurement and the ATLAS $p_T(t)$ distribution.
In Figure~\ref{fig:allexclusions} we show the region excluded by direct searches 
for $\phi_T$ and the exclusion curves from indirect searches for $y_{DM} =5$ and $y_{DM} = 10$. The regions to the left of the blue (red) curves are excluded at 95\% C.L. by the CMS invariant mass (ATLAS $p_T$) measurement. 
We also show the corresponding  curves obtained assuming the EFT approximation.
As expected from the results in the previous Section, the $p_T(t)$ measurement is more sensitive to the BSM signal and excludes masses up to $m_T \simeq 630$~GeV in the highly compressed region, if we take $y_{DM} = 5$.
This exclusion goes slightly beyond the masses probed by direct searches in the compressed scenario and illustrate the complementarity between the two types of searches.
For the same value of the BSM coupling, $y_{DM} = 5$, the exclusion curves obtained using the invariant mass distribution fall inside the direct search excluded region and are not shown.
Once we consider $y_{DM} = 10$, the $m(t\bar{t})$ measurement becomes competitive with direct searches in the compressed region and exclude masses up to $m_T \simeq 800$~GeV.
But the limits obtained from the $p_T(t)$ distribution are still stronger, excluding up to $m_T \simeq 1.1$~TeV and are complementary to direct searches even beyond the compressed region.

In Figure~\ref{fig:allexclusions} we also display the exclusion curves obtained assuming the EFT approximation.
As discussed in Sec.~\ref{sec:differential}, the EFT regime is not valid for the 
range of BSM masses considered here and the energies probed by the top measurements. Indeed we see that the EFT calculation considerably underestimates the excluded regions.
In particular, for $y_{DM} = 5$, the region excluded by top measurements falls completely inside the region already excluded by direct searches if we assume the EFT distributions.

Note that if we assume this minimal model fully explains the Dark Matter relic abundance, values of $y_{DM} \gtrsim 3$ would likely be excluded by DM Direct Detection searches~\cite{Garny:2018icg}. In this case, the $t\bar{t}$ measurements discussed here are not yet competitive to other searches. Nonetheless, once more LHC data is collected, the constraints obtained using the full 1-loop calculations could become relevant. The same is not true for the EFT analysis, since its constraints are too conservative. Therefore considering the 1-loop results is indeed essential for properly assessing the impact of top measurements to Dark Matter models.

All the results presented in Fig.\ref{fig:allexclusions} rely on $t\bar{t}$ distributions and do not exploit the full range of LHC and low energy data. On the other hand, global fits of SMEFT operators include a large number of observables.
Although the validity of SMEFT is limited to the high mass region,  it is still interesting to compare the results from the global fit in Ref.~\cite{Ellis:2020unq} and shown in Table~\ref{tab:limits} with the limits from Fig.\ref{fig:allexclusions}. First we point out that the strongest limit from the global fit for a single operator gives $|C_{tG}|/\Lambda^2 =|C_g| < 0.1$~TeV$^{-2}$, translating to $m_T \gtrsim$ 400 GeV for $y_{DM} = 10$, while the corresponding constraint we have obtained using the EFT approximation is $m_T \gtrsim$ 800 GeV.   Note, though, that by considering a single Wilson coefficient in the SMEFT analysis we are not making use of the correlations between the gluon-top $\Op{tG}$ and four-fermion $\Opp{tX}{(8)}$ operators, which explains the weaker bound.
Second, we have shown that the EFT result tends to underestimate the constraints. Therefore these two factors considerably enhances the SM measurements sensitivity to new physics when compared to a global SMEFT analysis.

\section{Conclusions}\label{sec:concls}

In this paper we have presented a new way to re-interpret SM top measurements which 
can access light new physics scales not suitable for the SMEFT framework, providing a path to go beyond the Top EFT approach.
In particular, we propose to consider new physics scenarios which produce loop-induced signatures in SM final states, motivated by the existence of Dark Matter.
In order to properly assess the sensitivity of top measurements to the BSM signal, we have computed the leading one-loop BSM contributions to top 
pair production through the use of form-factor effective couplings, which fully capture the BSM one-loop effects and can differ significantly from the behavior of SMEFT operators.
We have found that,  while the use of Top EFT operators produce an excess in the tail of distributions, the behavior of the loop calculation can be similar to a very broad bump.
Therefore, in order to fully capture the effects of new physics
one cannot simply re-interpret current SMEFT analyses, but should extend them making use of different kinematic regions other than distribution tails.
Our work  shows that this re-interpretation is possible and would make use of the same differential measurements considered by the experimental collaborations, just focusing in a different region of phase space.
This procedure could be generalized to other SMEFT sectors, like diboson and Higgs observables.

In addition, using the information of the full form-factor, SM measurements can be more sensitive to new physics than expected by the usual SMEFT analysis. 
We have shown that, for sufficiently large BSM couplings,
this enhancement in sensitivity renders the SM measurements complementary to direct searches, allowing to extend the excluded region of the BSM parameter space.

We point out that the results presented here are mostly intended to illustrate the potential of considering the one-loop form factors when modeling new physics contributions to SM measurements as opposite to the standard EFT approach.
A precise determination of the constraints from indirect searches would
require the inclusion of other top measurements, the calculation of QCD NLO corrections to the signal and more detailed information about the SM uncertainties and their correlations, which we leave for future work.
Finally, the broad resonant behavior of the signal and the negative interference at high energy bins could also be better exploited to constrain new physics. In particular, ratios of intermediate and high invariant mass bins could in principle reduce the systematical uncertainties and enhance the signal sensitivity.

\acknowledgments

We would like to thank Maeve Madigan for her help with the reinterpretation of the ATLAS and CMS top analyses and Johnathon Gargalionis for his help with the EFT matching procedure. 
A.L.\ is supported by FAPESP grants no. 2018/25225-9 and 2021/01089-1. 
The research of VS is supported by the Generalitat
Valenciana PROMETEO/2021/083 and the Ministerio de Ciencia e  Innovacion PID2020-113644GB-I00.

\appendix

\include{appendix}

\newpage

\bibliographystyle{JHEP}
\bibliography{references}
\end{document}

%% file: appendix.tex
\section{Feynman Diagrams} \label{app:diagrams}

\begin{figure}[h!]
  \includegraphics[scale=1]{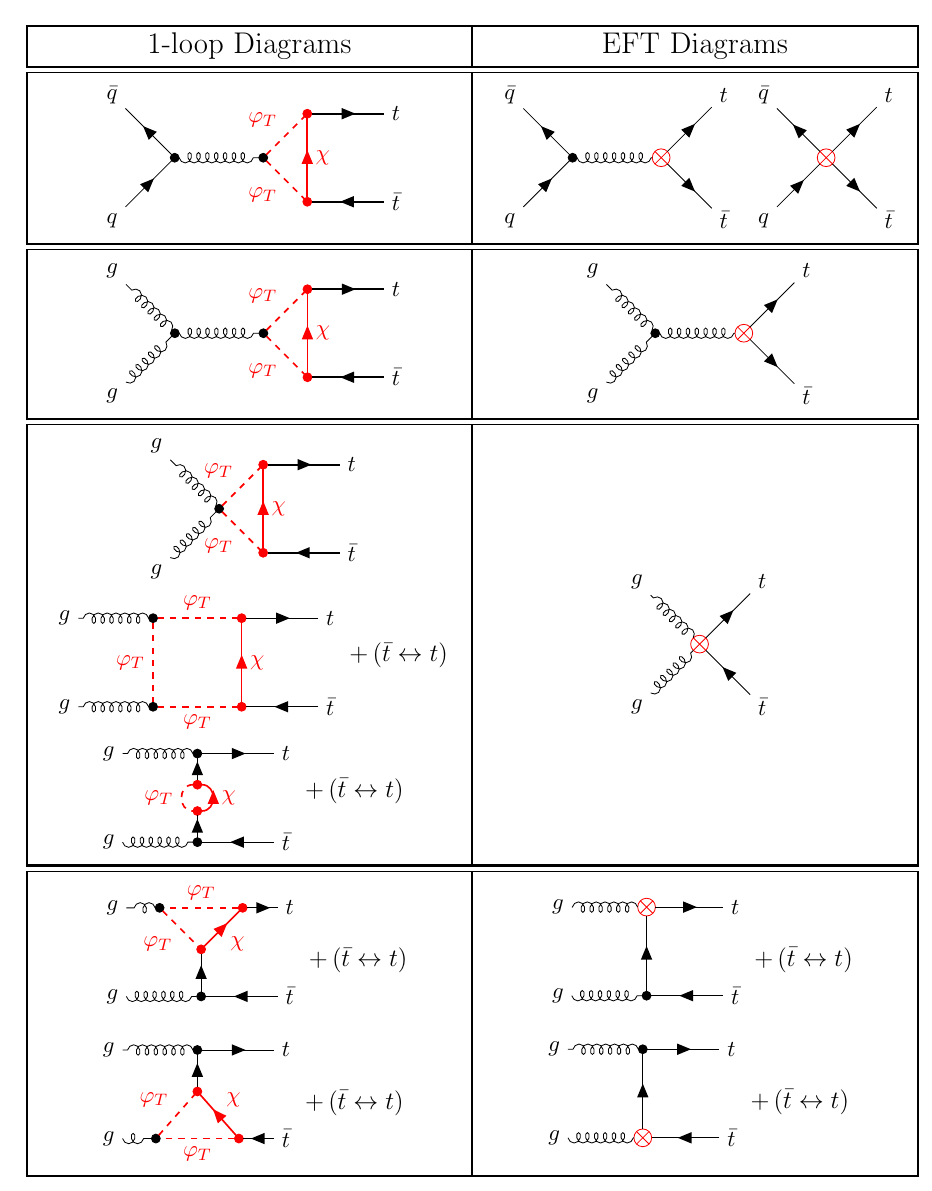}
\caption{Loop diagrams and their EFT equivalents contributing to $\bar{t} t$ production at leading order in the BSM coupling $y_{DM}$. \label{tab:diagrams}}
\end{figure}

In this Section we list all the relevant Feynman diagrams required for computing the BSM contributions to $t\bar{t}$ production at the LHC at leading order in the BSM coupling $y_{DM}$.
The diagrams are shown in Figure~\ref{tab:diagrams} and are divided into quark initiated (first row) and gluon initiated (second, third and fourth rows) processes. In the left column we show the diagrams used for the 1-loop calculation, while the right column shows the equivalent ones used within the EFT approximation. The diagrams which correspond to a top/anti-top permutation of other diagrams are not explicitly shown, but are indicated by $\left( \bar{t} \leftrightarrow t \right)$.
The counter-terms diagrams needed for regularizing the divergent diagrams are not shown.

\section{Form Factors and EFT matching}
\label{app:formFactors}

In order to verify the expression for the form factor $\mathcal{F}^\mu$ defined in Eq.(\ref{eq:fmu}) and the EFT coefficients $C_g,C_q$ defined in Eqs.(\ref{eq:Cq}) and (\ref{eq:Cq}), we compute the amplitude for the quark initiated process, $q \bar{q} \to t \bar{t}$,  using both the 1-loop form factors and the EFT operators. The corresponding diagrams are shown by the first row in Figure~\ref{tab:diagrams}. In the EFT limit, $m_T,m_\chi \gg m_t,\sqrt{s}$, we should obtain the same result for both approaches, thus validating our implementation. A similar exercise could in principle be done for the $g g \to t \bar{t}$ process. However, due to the large number of diagrams (see Figure~\ref{tab:diagrams}) and the complexity of the form factor, the validation of $\mathcal{F}^{\mu\nu}$   has only been done numerically.

We start with the 1-loop calculation using form factors. Note that for the $q \bar{q} \to t \bar{t}$ process only the top-top-gluon form factor ($\mathcal{F}^\mu$) is relevant and the top momenta appearing in $\mathcal{F}^\mu$ are on-shell, $p_1^2 = p_2^2 = m_t^2$. In this case the form factor defined in Eq.(\ref{eq:fmu}) can be simplified, since $C_{1} = C_{1}(p_1^2,s,p_2^2) = C_{1}(m_t^2,s,m_t^2) = C_{2}$ and similarly $C_{22} = C_{11}$. Using these results we obtain:
\begin{align}
	\mathcal{F}^\mu = & i T^A P_R  \left[ \left( C_1 + 2 C_{11} \right) \left( \slashed{p}_1 p_1^\mu + \slashed{p}_2 p_2^\mu\right)  - \left( C_1 + 2 C_{12} \right) \left( \slashed{p}_1 p_2^\mu + \slashed{p}_2 p_1^\mu\right) \right] \nonumber \\ 
	& + 2 i T^A P_R   \gamma^\mu \left(C_{00} + \delta_{R} \right) + 2 i T^A P_L \gamma^\mu \delta_{L} \label{eq:fmuOn}
\end{align}
where $C_{i}, C_{ij}$ are the triangular scalar loop functions and $\delta_{L,R}$ are the counter-terms. Using NLOCT~\cite{Degrande:2014vpa}, dimensional regularization and the on-shell renormalization conditions we obtain\footnote{The expressions in Eq.(\ref{eq:CTs}) for the counter-term are valid for $m_T > m_\chi + m_t$.  Similar expressions can be found for the compressed region and have been used in our results for scenarios with $m_T < m_\chi + m_t$.}:
\begin{align}
	\delta_L = & \frac{1}{128 \pi ^4 x_t^2 \sqrt{\lambda }} \biggl\{  2 \sqrt{\lambda } \left[ x_t \left(x_t+2 x+\log (x)-2\right)-(x-1)^2 \log (x)\right]  \nonumber\\
		& \qquad  -4 \left[ x_t \left(x_t+x^2+x-2\right)-(x-1)^3\right] \log \left(\frac{\sqrt{\lambda }-x_t+x+1}{2 \sqrt{x}}\right) \biggr\} \nonumber \\
	\delta_R = & \frac{\left(x-1-x_t\right)}{128 \pi ^4 x_t^2 \sqrt{\lambda}} \biggl\{  2	\left[ \left(x_t-1\right)^2 + (x-2) x \right] \log\left(\frac{\sqrt{\lambda }-x_t+x+1}{2\sqrt{x}}\right) \nonumber \\
	& \qquad +\sqrt{\lambda } \left[ 2x_t-\left(x_t+x-1\right) \log(x)\right] \biggr\}  - \frac{1}{64 \pi^4}\left[\frac{1}{\epsilon} + \log\left(\frac{\mu^2}{m_T^2}\right)\right]
	 \label{eq:CTs}
 \end{align}
where $x = m_\chi^2/m_T^2, x_t = m_t^2/m_T^2$ and $\lambda = 1+x^2+x_t^2-2 x x_t-2 x_t-2 x$. Both the divergent ($1/\epsilon$) and the $\log(\mu^2/m_T^2)$ terms in $\delta_R$ cancel with the corresponding terms from $C_{00}$, which is the only divergent loop integral in Eq.(\ref{eq:fmuOn}).

Using the results above, the amplitude for the first diagram in Figure~\ref{tab:diagrams} becomes:
\begin{align}
	\mathcal{M}_{\rm loop} = & -i \pi^2 y_{DM}^2 g_s^2 \frac{1}{s} T^A \SPqq{\gamma^\mu} \SPtt{\left(-i \mathcal{F}_\mu \right)} \nonumber \\
	= &  -i \pi^2 y_{DM}^2 g_s^2 \frac{1}{s} T^A \, T^A \biggl\{  2 \SPqq{\gamma^\mu} \left[ \left( C_{00} + \delta_R \right) \SPtt{\gamma_\mu P_R} + \delta_L \SPtt{\gamma_\mu P_L}  \right] \nonumber \\
	&\quad  + \SPqq{\slashed{p}_1} \left[ \left(C_1 + 2 C_{11} \right) \SPtt{\slashed{p}_1 P_R} - \left( C_1 + 2 C_{12} \right) \SPtt{\slashed{p}_2 P_R} \right] \nonumber \\
	& \quad   + \SPqq{\slashed{p}_2} \left[ \left(C_1 + 2 C_{11} \right) \SPtt{\slashed{p}_2 P_R} - \left( C_1 + 2 C_{12} \right) \SPtt{\slashed{p}_1 P_R} \right] \biggr\}	\label{eq:ampLoop1}
\end{align}
where $u_{q(t)}$ is the quark (top) spinor and $v_{\bar{q} (\bar{t})}$ is the anti-quark (anti-top) spinor. The color indices for the Gell-Mann matrices $T^A$ are contracted with the spinors and are not explicitly shown for simplicity.

The amplitude can be greatly simplified using the on-shell relations: 
\begin{align}
	\slashed{p} u_{p} = m_t u_p, & \quad \bar{u}_p \slashed{p} = m_t \bar{u}_p \nonumber \\
	\slashed{p} v_{p} = -m_t v_p, & \quad \bar{v}_p \slashed{p} = -m_t \bar{v}_p
\end{align}
and $\slashed{p} P_R = P_L \slashed{p}$. Applying these relations to Eq.(\ref{eq:ampLoop1}) we obtain:
\begin{align}
	\mathcal{M}_{\rm loop} = &  -i \pi^2 y_{DM}^2 g_s^2 \frac{1}{s} T^A \, T^A \biggl\{ 2 \SPqq{\gamma^\mu} \left[ \left( C_{00} + \delta_R \right) \SPtt{\gamma_\mu P_R} + \delta_L \SPtt{\gamma_\mu P_L}  \right]  \nonumber \\
	& \qquad + m_t \left( C_{11} - C_{12} \right) \SPqq{\left( \slashed{p}_1 + \slashed{p}_2 \right)} \SPtt{\left(P_R-P_L\right)} \nonumber \\
	& \qquad   + m_t \left( C_1 + C_{11} + C_{12} \right) \SPqq{\left( \slashed{p}_1 - \slashed{p}_2 \right)} \SPtt{} \biggr\}	\label{eq:ampLoop2}
\end{align}

Finally, using momentum conservation, $\slashed{p}_1 + \slashed{p}_2 =\slashed{k}_1 + \slashed{k}_2$, where $k_{1,2}$ are the initial state momenta and the on-shell relations for the massless incoming quarks we have: $\SPqq{\left( \slashed{p}_1 + \slashed{p}_2 \right)} = 0$. Hence:
\begin{align}
	\mathcal{M}_{\rm loop} = & -i \pi^2 y_{DM}^2 g_s^2 \frac{1}{s} T^A \, T^A \biggl\{ \nonumber\\ 
	& \SPqq{\gamma^\mu} \left[ 2 \left( C_{00} + \delta_R  - \delta_L \right) \SPtt{\gamma_\mu P_R} + 2 \delta_L \SPtt{\gamma_\mu}  \right] \nonumber \\
	&   + m_t \left( C_1 + C_{11} + C_{12} \right) \SPqq{\left( \slashed{p}_1 - \slashed{p}_2 \right)} \SPtt{} \biggr\}	\label{eq:ampLoop3}
\end{align}
where we have used $P_L = 1-P_R$.

The above result corresponds to the full 1-loop calculation for the $q \bar{q} \to t \bar{t}$ process. In order to compare it to the EFT approximation, we compute the same process, but now using the EFT lagrangian defined in Eq.(\ref{eq:lEFTonshell}) and the diagrams shown in the first row (right column) of Figure~\ref{tab:diagrams}.
The EFT amplitude in this case is simply:
\begin{align}
	\mathcal{M}_{\rm EFT} = & - 2 i g_s  \frac{m_t}{s}  T^A \, T^A C_g \left[  2 m_t \left(\SPqq{\gamma^\mu}\right) \left(\SPtt{\gamma_\mu}\right) +  \SPqq{ \left(\slashed{p}_2 - \slashed{p}_1\right)}  \left( \SPtt{} \right)    \right]  \nonumber \\
	& - i T^A \, T^A C_q \left( \SPqq{\gamma^\mu} \right) \left( \SPtt{\gamma_\mu P_R} \right) \label{eq:ampEFT1}
\end{align}
where $C_q,C_g$ are the EFT coefficients defined in Eqs.(\ref{eq:Cq}) and (\ref{eq:Cq}). Before we can compare the EFT result with the 1-loop calculation it is useful to write the above amplitude as:
\begin{align}
	\mathcal{M}_{\rm EFT} = & -i \pi^2 y_{DM}^2 g_s^2 \frac{1}{s} T^A \, T^A \biggl\{ \vphantom{\left[\left(\slashed{p}_1\right)\right]} \nonumber\\ 
	&  \SPqq{\gamma^\mu} \left[ 2 \left(\frac{s C_q}{2 \pi^2 g_s^2 y_{DM}^2}\right)  \SPtt{\gamma_\mu P_R} + 2 \left(\frac{2 m_t^2 C_g }{\pi^2 g_s y_{DM}^2}\right) \SPtt{\gamma_\mu} \right]   \nonumber \\
	&  + m_t \left( \frac{-2 C_g}{\pi^2 g_s y_{DM}^2} \right) \SPqq{\left( \slashed{p}_1 - \slashed{p}_2 \right)} \SPtt{} \biggr\} \label{eq:ampEFT2}
\end{align}

Comparing Eqs.(\ref{eq:ampLoop3}) and (\ref{eq:ampEFT2}) we see that, {\it in the EFT limit}, we must have:
\begin{align}
	C_1 + C_{11} + C_{12} = & \left(\frac{-2 C_g}{\pi^2 y_{DM}^2 g_s}  \right) \nonumber \\
	C_{00} + \delta_R - \delta_L = & \left(\frac{s C_q}{2 \pi^2 g_s^2 y_{DM}^2}  \right) \nonumber\\
	\delta_L = & \left(\frac{2 m_t^2 C_g}{\pi^2 y_{DM}^2 g_s}\right) \label{eq:matchingEqs}
\end{align}

In order to verify the above relations we must expand the loop functions $C_i,C_{ij}$ and the counter-terms $\delta_{L,R}$ to leading order in $s,m_t$, which gives:
\begin{align}
	& C_1 + C_{11} + C_{12} = \frac{1}{192 \pi ^4}\frac{1}{m_T^2}\frac{1}{(x-1)^4} \left[1-6x+3x^2+2x^3-6x^2\log(x)\right] + \mathcal{O}(\frac{1}{m_T^4}) \nonumber \\
	& C_{00} + \delta_R - \delta_L = \frac{1}{1152 \pi ^4}\frac{s}{m_T^2}\frac{1}{(x-1)^4} \left[2-9x+18x^2-11x^3+6x^3\log(x) \right]  + \mathcal{O}(\frac{1}{m_T^4}) \nonumber\\
	& \delta_L = \frac{-1}{192 \pi ^4}\frac{m_t^2}{m_T^2}\frac{1}{(x-1)^4} \left[1-6x+3x^2+2x^3-6x^2\log(x)\right] + \mathcal{O}(\frac{1}{m_T^4}) 
	\label{eq:coeffEFT}
\end{align}

Finally, comparing the above results with the expressions for $C_q$ and $C_g$ given in Eqs.(\ref{eq:Cg}) and (\ref{eq:Cq}), we see that the matching relations in Eq.(\ref{eq:matchingEqs}) are indeed satisfied, as expected.

\section{Indirect Searches - Expected Limits}
\label{app:expectedLimits}

\begin{figure}[!h]
    \centering
    \includegraphics[width=0.95\textwidth]{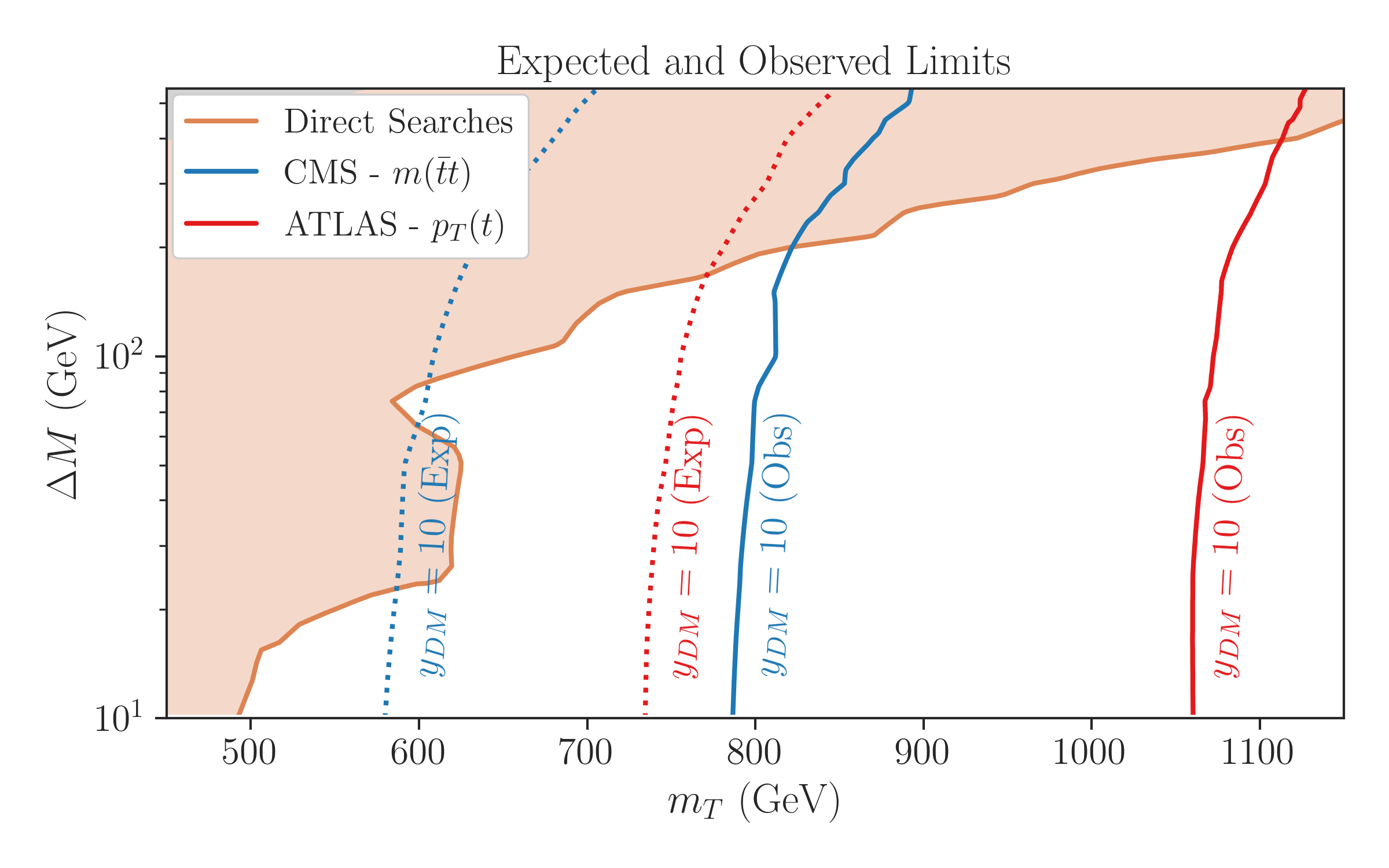}
    \caption{Comparison of the exclusion curves from direct stop searches and from top measurements. The solid (dotted) curves show the observed (expected) exclusions. All the limits were computed using the 1-loop form factors with $y_{DM} = 10$. The limits obtained using the CMS invariant mass measurement from Ref.\cite{CMS-TOP-20-001} are shown by the blue curves, while the ones obtained from the ATLAS $p_T$ measurement\cite{ATLAS-TOPQ-2019-23} are shown in red.}
    \label{fig:allexclusionsExp}
\end{figure}

All the indirect searches constraints presented in Sec.\ref{sec:results} make use of the unfolded top measurements discussed in Sec.\ref{sec:indirectSearches}.
As shown in Figs.\ref{fig:cmsDists} and \ref{fig:atlasDists}, both the invariant mass and transverse momentum distributions display a few bins where the SM prediction is above the measured values. Although these differences are within two standard deviations\cite{CMS-TOP-20-001}, they result in stronger limits than expected. In order to illustrate and quantify this difference we compute the {\it expected} limits on the signal using the CMS $m(t\bar{t})$ and ATLAS $p_T(t)$ measurements following the procedure outlined in Sec.\ref{sec:indirectSearches}, but now assuming $N^i_{\rm Obs} = N^i_{\rm SM}(\rm NNLO)$. The results are shown in Figure~\ref{fig:allexclusionsExp}, where we display the exclusion curves at 95\% C.L. assuming $y_{DM} = 10$ and using the 1-loop calculation. As we can see the exclusion on $m_T$ is reduced by $\sim 200-300$~GeV when we compare the observed and expected exclusions.
A better treatment of the SM predictions and their uncertainties will likely bring the observed results within better agreement with the expected exclusion.
Nonetheless, our overall conclusions about the potential complementarity between direct and indirect searches and the lack of validity of the EFT approximation still hold.

